\documentclass[twocolumn,10pt]{article}
\usepackage{fullpage}

\usepackage{amsfonts,amssymb}
\usepackage{booktabs}
\usepackage{adjustbox}
\usepackage{hyperref}

\usepackage{mathtools}
\usepackage{multirow}

\DeclarePairedDelimiterX{\norm}[1]{\lVert}{\rVert}{#1}
\DeclarePairedDelimiterX{\abs}[1]{\lvert}{\rvert}{#1}

\makeatletter
\newcommand*\bigcdot{\mathpalette\bigcdot@{.5}}

\DeclareMathOperator*{\E}{\mathbb{E}}

\newcommand{\smoothTerm}{s}

\newcommand{\average}{\mu}
\newcommand{\populationSymbol}{P}
\newcommand{\population}[2]{P_{#1 #2}}
\newcommand{\exports}[3]{X_{#1 #2 #3}}
\newcommand{\RCAfraction}[3]{\frac{\exports{c}{p}{t} / \sum_p \exports{c}{p}{t} }{\sum_c \exports{c}{p}{t} / \sum_{cp} \exports{c}{p}{t} }}
\newcommand{\RCAsymbol}[3]{\textrm{RCA}_{#1 #2 #3}}
\newcommand{\nullExportsSymbol}[3]{\E \left [ \exports{#1}{#2}{#3} \vert \population{c}{t} \right ]}
\newcommand{\nullGlobalExportsSymbol}[2]{\E \left [\sum_c  \exports{c}{#1}{#2} \Bigg \vert \sum_c  \population{c}{t} \right ]}
\newcommand{\numProducts}{59}
\newcommand{\productCode}[1]{\texttt{#1}}
\newcommand{\stddev}{\sigma}
\newcommand{\products}{\mathcal{P}}
\newcommand{\countries}{\mathcal{C}}

\newcommand{\countrySpecificIntercept}[1]{\alpha_{#1}}
\newcommand{\globalIntercept}[1]{\gamma_{#1}}
\newcommand{\countrySpecificExponent}[1]{\beta_{#1}}
\newcommand{\globalExponent}[1]{\delta_{#1}}

\newcommand{\nullExports}[1]{
		\countrySpecificIntercept{#1}
		\left (
			\population{c}{t}
		\right )^{\countrySpecificExponent{#1}}
}
\newcommand{\nullGlobalExports}[1]{
		\globalIntercept{p}
		\left (
			\sum_{c} \population{c}{t}
		\right )^{\globalExponent{p}}
}

\newcommand{\RpopFraction}[3]{
\frac{
	\exports{c}{p}{t}
	/
	\left (
		\nullExports{p}
	\right )
}
{
	\sum_c \exports{c}{p}{t}
	/
	\left (
		\nullGlobalExports{p}
	\right )
}
}
\newcommand{\RpopSymbol}[3]{\mathcal{R}_{#1 #2 #3}}
\newcommand{\RpopSymbolCenteredScaled}[3]{{R}_{#1 #2 #3}}

\newcommand{\Xminpos}{x_m}
\newcommand{\diversity}[2]{\texttt{diversity}_{#1 #2}}

\newcommand{\countriesToRemove}{\texttt{CountriesToRemove}}
\newcommand{\productsToRemove}{\texttt{ProductsToRemove}}

\newcommand{\scorePC}[1]{\phi_{#1}}
\newcommand{\scoreFirstPC}{\scorePC{0}}
\newcommand{\scoreSecondPC}{\scorePC{1}}
\newcommand{\gdppc}{\texttt{GDPpc}}
\newcommand{\link}[1]{g \left ( #1 \right )}
\newcommand{\linkExpression}[1]{\text{sign}(x) \abs{x}^\linkRoot}
\newcommand{\linkRoot}{{1/2}}
\newcommand{\beginningYearTrainingSet}{{1962}}
\newcommand{\finalYearTrainingSet}{{1988}}

\newcommand{\method}{Principal Smooth-Dynamics Analysis}
\newcommand{\methodAcronym}{PriSDA}

\newcommand{\gamIntercept}[1]{c_{#1}}

\newcommand{\paperTitle}{Machine-learned patterns suggest that diversification drives economic development}

\usepackage{graphicx}
\makeatletter
\let\saved@includegraphics\includegraphics
\AtBeginDocument{\let\includegraphics\saved@includegraphics}
\renewenvironment*{figure}{\@float{figure}}{\end@float}
\makeatother

\usepackage{authblk}

\begin{document}
\title{\paperTitle}

\author[1]{Charles D.\ Brummitt}
\author[2]{Andr\'es G\'omez-Li\'evano}
\author[2,3]{Ricardo Hausmann}
\author[1]{Matthew H. Bonds}
\affil[1]{Global Health and Social Medicine at Harvard Medical School}
\affil[2]{Center for International Development at Harvard University}
\affil[3]{Harvard Kennedy School}

\twocolumn[
  \begin{@twocolumnfalse}
  \maketitle
\begin{abstract}
We develop a machine-learning-based method, \method\ (\methodAcronym), to identify patterns in economic development and to automate the development of new theory of economic dynamics. Traditionally, economic growth is
modeled with a few aggregate quantities derived from simplified theoretical models. 
Here, \methodAcronym\ identifies important quantities. 
Applied to 55 years of data on countries' exports, 
 \methodAcronym\ finds that what most distinguishes countries' export baskets is their diversity, 
with extra weight assigned to more sophisticated products. 
The weights are consistent with previous measures of product complexity in the literature. The second dimension of variation is a proficiency in machinery relative to agriculture.  
\methodAcronym\ then couples these quantities with per-capita income and infers the dynamics of the system over time. 
According to \methodAcronym, the pattern of economic development of countries is dominated by a tendency toward increased diversification. 
Moreover, economies appear to become richer after they diversify (i.e., diversity precedes growth). 
The model predicts that middle-income countries with diverse export baskets will grow the fastest in the coming decades, and that countries will converge onto intermediate levels of income and specialization. 
\methodAcronym\ is generalizable and 
may illuminate dynamics of elusive quantities such as diversity and complexity in other natural and social systems.
\end{abstract}  \end{@twocolumnfalse}
]


Computers can learn to predict the future using vast datasets too large for humans to grasp. 
The logic behind a machine's prediction, however, is often inscrutable\ \cite{Voosen2017}, and accuracy, not insight, is often the goal. 
Scientists, meanwhile, generate new theories using intuition or derive them incrementally from existing models. 
Recent advances in interpretable machine learning are enabling the generation of theories to be automated:
machines have identified mathematical laws in physical and biological systems 
that took many years for scientists to solve manually\ \cite{Bongard2007,Schmidt2009,Daniels2015,Zhang2015,Brunton2016_SINDy}. 
But while pendulums and fluid dynamics
may follow elegant governing equations, systems comprising capricious humans may not. Thus, in social sciences it is challenging for machines to teach humans better theories.

We introduce a method for identifying interpretable patterns in high-dimensional times-series called \emph{\method} (\methodAcronym), and apply it to a unifying question within the social sciences: why some countries are rich and others are poor. \methodAcronym\ ingests time-series data from a system characterized by potentially many dimensions, and it identifies a small number of dynamical equations involving smoothing splines that model how the system changes over time. The resulting model is optimized for accuracy yet is readily interpretable. 

Traditionally, one builds an economic theory by manually choosing a dimension-reduced representation of a complex process, such as ``utility'' or an aggregated value such as gross domestic product. Then one constructs models of how those few quantities change over time, and finally one estimates the derived model with data. These models often neglect important patterns, such as the emergence of the tremendous diversity of goods or why rich countries trade with other rich countries \cite{Krugman1993_Geography_and_Trade}. Agnostic of explicit traditions of economics, what theory of economic growth would an automated economist find?

Here, we analyze machine-reduced dimensions of countries' export baskets (corrected for population size) using principal components analysis\ \cite{Lever2017}, and then we train generalized additive models (GAMs)\ \cite{Wood2006book} to predict yearly changes in export baskets and in per-capita income. The results indicate the fundamental importance of an economy's (complexity-weighted) diversity, both as a summary statistic---diversity captures more variance of export baskets across time than any other direction---and as a predictor of growth. The method, \methodAcronym, is generally applicable and may illuminate the emergence of diversity and complexity in other biological, physical, and social systems.

\begin{figure*}[htb]
\begin{center}
\includegraphics{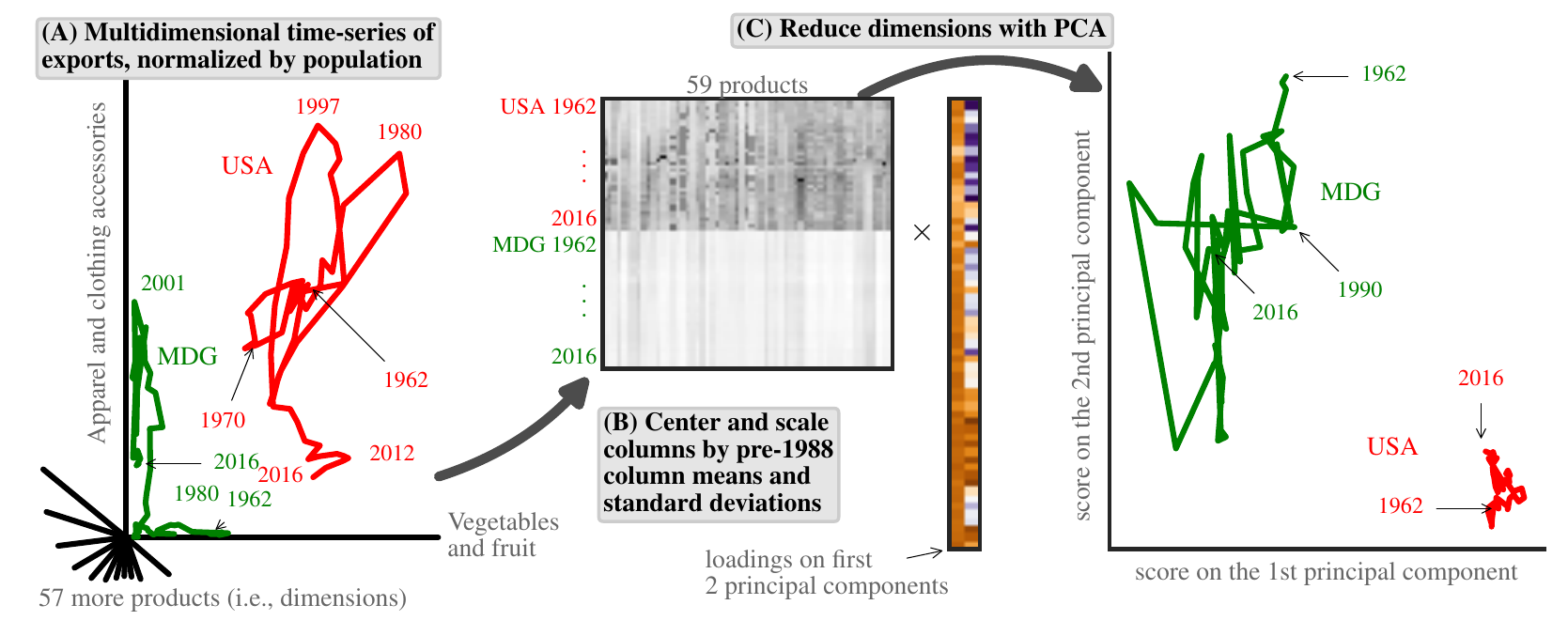}
\caption{
\textbf{Preprocessing and reducing dimensions of export baskets.}
(A) 
Begin with time-series of export values of $\numProducts$ product categories, normalized by population 
[Eq.\ (SI-1)]
and logarithmically transformed 
[Eq.\ (SI-5)]
to make large and small countries comparable. 
For illustration, we show the trajectories of the United States (USA) and Madagascar (MDG). 
We are illustrating two dimensions here, but in reality the red and green curves live in $\numProducts$ dimensions.
(B) Center and scale columns by their pre-$\finalYearTrainingSet$ means and standard deviations.
(C) Reduce dimensions with principal components analysis (PCA). 
Each country's score in a given principal component represents a certain linear combination of its export basket. Together, the scores on the first few principal components summarize the country's export basket with just a few numbers. 
}
\label{fig:overview}
\end{center}
\end{figure*}

\section{Methods}
%

The \methodAcronym\ method identifies equations that predict changes over time (using nonparametric regression, here GAMs) in a complex system described using dimension reduction and potentially other aggregate quantities. Crucial to its success is providing it data that allows large and small units (here, national economies) to be comparable yet allow for absolute growth.

\subsection{Multidimensional characterization of the productive capabilities of economies\label{sec:define_absolute_advantage}}
There is growing interest in multi-dimensional metrics of economic development and poverty\ \cite{Alkire2011,Hruschka2017}.  
We track macro-level multidimensional  economic development based on annual exports, for which there is high quality data for all countries over the past half century. 
A country's exports indicate its international competitive advantages, and unlike domestic production, exports share a common classification system. 

Instructions for accessing the data and details on its preprocessing are in 
Sections SI-2 and SI-3.
Because data on exports are noisy, products are aggregated into $\numProducts$ categories 
(Section SI-3A).
The value of a country $c$'s exports of a product $p$ in year $t$, denoted $\exports{c}{p}{t}$, tends to correlate positively with the size of the country's population, $\population{c}{t}$. To account for that relationship, $\exports{c}{p}{t}$ is divided by an expectation according to a null model of a country's expected value of an export {given} that country's population. 
To remove the effects of global price shocks, this quantity is divided by the total value of the world's exports of that product, which is also normalized by a null model that predicts global export value using global population. 
We call the resulting quantity the \emph{absolute advantage} of a country $c$ in a certain product $p$ in year $t$, denoted $\RpopSymbol{c}{p}{t}$:
\begin{align}
\RpopSymbol{c}{p}{t} =
\frac
{
	\exports{c}{p}{t} / \E \left [ \exports{c}{p}{t} \vert \population{c}{t} \right ]
}
{
	\sum_c \exports{c}{p}{t} / \E \left [\sum_c  \exports{c}{p}{t} \big \vert \sum_c  \population{c}{t} \right ]
}
.
\label{eq:define_absolute_advantage}
\end{align}
Details are in Sec.\ SI-3B. 
We consider countries as length-$\numProducts$ vectors of $\RpopSymbol{c}{p}{t}$ across all products; a two-dimensional projection of two trajectories of $\RpopSymbol{c}{p}{t}$ is shown in Fig.\ \ref{fig:overview}(A). Unlike relative quantities such as revealed comparative advantage\ \cite{Balassa1965}, a country can grow its absolute advantage arbitrarily. 
For example, in 2016 Belgium and the Netherlands were the only countries that ``punched above their weight'' (i.e., had $\RpopSymbol{c}{p}{t} > 1$) for all $\numProducts$ products $p$. 

To put products on equal footing with each other, we center and scale $\RpopSymbol{c}{p}{t}$ by the mean and standard deviation of $\RpopSymbol{c}{p}{t}$ across all countries and across all years $t \leq \finalYearTrainingSet$ (Fig.\ \ref{fig:overview}(B)).
We call the resulting quantity \emph{scaled absolute advantage} and denote it by $\RpopSymbolCenteredScaled{c}{p}{t}$. 
Scaled absolute advantage is the number of standard deviations above the pre-$\finalYearTrainingSet$ mean of all countries' absolute advantage in that product. $\RpopSymbolCenteredScaled{c}{p}{t} > 0$ means that country $c$ excels at producing and exporting product $p$ in year $t$. Making products comparable---by dividing by the product's global export market in $\RpopSymbol{c}{p}{t}$ and by centering and scaling each product in $\RpopSymbolCenteredScaled{c}{p}{t}$---enables detecting how expertise in one product enables developing expertise in another, regardless of the sizes of the markets of those products.

\subsection{Reducing dimensions of export baskets}

We reduce dimensions using principal components analysis (PCA)\ \cite{Lever2017} because the resulting dimensions are interpretable and because summing exports reduces the noise in export data. 
Other methods are discussed in Sec.\ SI-1.

\begin{figure*}[htb]
\begin{center}
\includegraphics{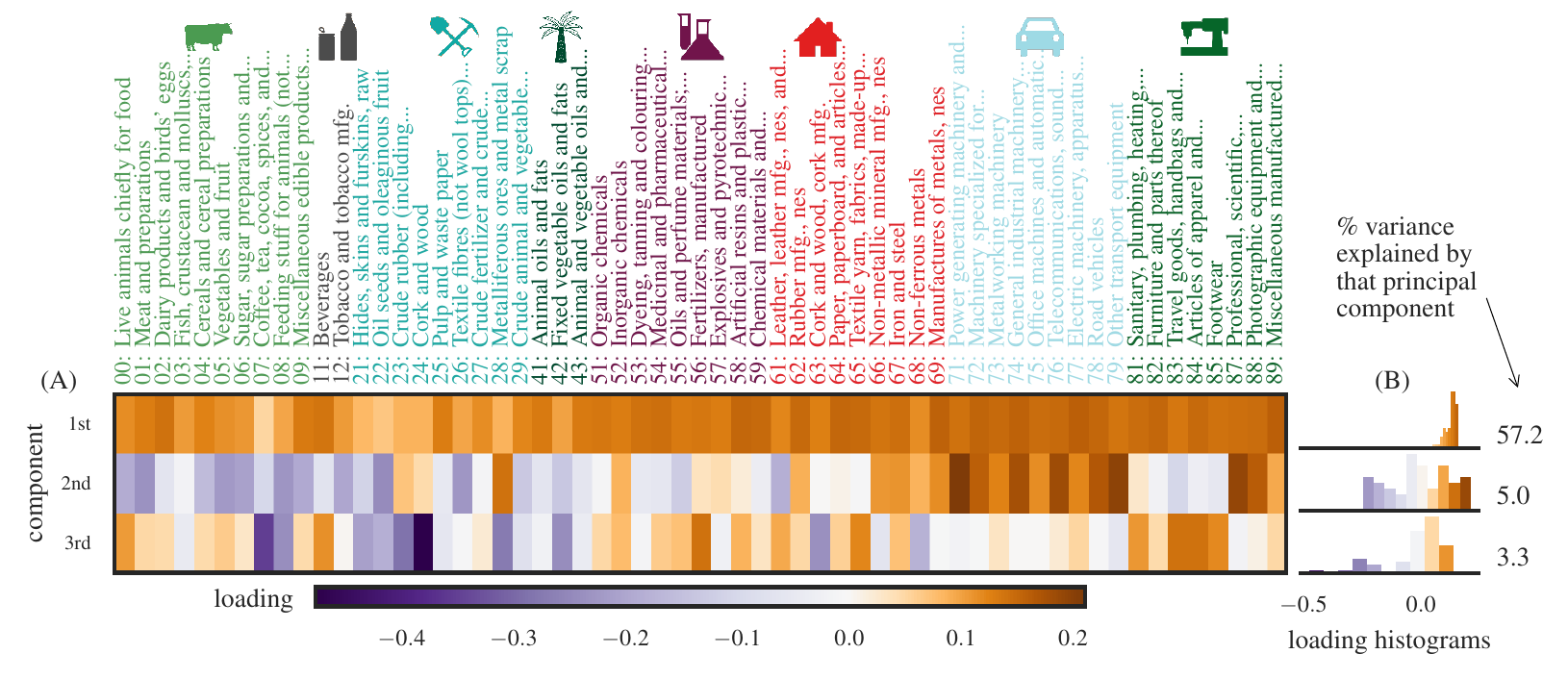}
\caption{
\textbf{The first three principal components are approximately
	(1) {total absolute advantage summed across all products (with more weight on product codes above \productCode{50})},
	(2) {machinery minus agriculture}, and
	(3) {textiles and fertilizer minus coffee and cork}.
}
In plot (A), the rows are principal components, the columns are products, and the rectangles' colors represent the loading (or ``weight'') of that principal component on that product.  
The first component loads positively on all products. 
Thus, what distinguishes countries, above all, is their ``diversification'' across products. 
The second component loads highly on machinery (product codes beginning with \productCode{7}) and other manufactured goods, and it loads negatively on agricultural products. 
Thus, the direction in $\numProducts$-dimensional product space orthogonal to the first component that most spreads out observations points toward machinery and away from agriculture. 
The third component loads positively on clothing and textile products and negatively on cork and wood (\productCode{24}) and coffee, tea, and spices (\productCode{07}). 
The plots labeled (B) are histograms of loadings, across all $\numProducts$ products, in the corresponding rows. 
}
\label{fig:pca_loadings}
\end{center}
\end{figure*}

Figure\ \ref{fig:pca_loadings} shows the \emph{loadings} (weights) of the first three principal components on the $\numProducts$ products. The \emph{score} of a country's export basket on the $k$th principal component---denoted $\scorePC{k}$---is the dot product [illustrated in Fig.\ \ref{fig:overview}(C)] of the country's export basket, $(\RpopSymbolCenteredScaled{c}{p}{t})_{p \in \products}$, with that principal component's loading vector, drawn as a row of rectangles in Fig.\ \ref{fig:pca_loadings}(A). We interpret this PCA in Sec.\ \ref{sec:results}.

\subsection{Inferring dynamics of export baskets}

To understand patterns in economic development, next we examine how two summary measures of an export basket, $\scoreFirstPC$ and $\scoreSecondPC$, interact with per-capita income\ \cite{GDPpcWorldBank} (transformed logarithmically):
$\gdppc \equiv \log_{10}(\text{GDP per capita})$.
Because excelling at exports reflects the capabilities and know-how within a country\ \cite{Hausmann2006,Hidalgo2009,Hausmann:2011ke}, by inferring a model of how the triple $(\scoreFirstPC, \scoreSecondPC, \gdppc)$ changes over time, we aim to shed light on fundamental economic patterns.

The three variables $(\scoreFirstPC, \scoreSecondPC, \gdppc)$ are aggregate descriptions of an economy, so we expect them to change smoothly over time. 
A natural choice for a smooth model are cubic smoothing splines\ \cite[Chapters 3 and 4]{Wood2006book}. This method provides us with the following system of dynamical equations: 
\begin{subequations}
\begin{align}
\link{\Delta \scoreFirstPC(t)} 
&= 
\gamIntercept{0}
+
\smoothTerm_{00} \left (\scoreFirstPC(t) \right )
+
\smoothTerm_{01} \left (\scoreSecondPC(t) \right ) 
+
\smoothTerm_{02} \left (\gdppc(t) \right ) \label{eq:pc0}  \\
\link{\Delta \scoreSecondPC(t)}
&= 
\gamIntercept{1}
+
\smoothTerm_{10} \left (\scoreFirstPC(t) \right ) 
+
\smoothTerm_{11} \left (\scoreSecondPC(t) \right ) 
+
\smoothTerm_{12} \left (\gdppc(t) \right ) 
\label{eq:pc1} 
\\
\link{\Delta \gdppc(t)}
&= 
\gamIntercept{2} + 
\smoothTerm_{20} \left (\scoreFirstPC(t) \right ) 
+
\smoothTerm_{21} \left (\scoreSecondPC(t) \right ) 
+
\smoothTerm_{22} \left (\gdppc(t) \right ) 
\label{eq:gdppc}
\end{align}
\label{eq:gam}
\end{subequations}
\!\!\!\!\!where $\Delta$ takes the expected difference in time, $\Delta f(t) \equiv \E \left [ f(t+1) - f(t) \right ]$; 
the $\gamIntercept{i}$ are intercept terms; 
the $\smoothTerm_{i, j}$ are cubic smoothing splines with smoothing strength chosen using nested-in-time cross validation 
(Sec.\ SI-5A); 
and the link function $\link{x} \equiv \linkExpression{x}$ is applied to make the residuals' distributions closer to a normal 
(Sec.\ SI-5B). 
The goodness of fit ($R^2 \approx 0.04$) and the GAM's competitiveness with other statistical learning methods are discussed in 
Sec.\ SI-5A.2.

The terms of \eqref{eq:gam} are plotted in Figure\ \ref{fig:partial_dependence}, where one can compare 
not only the shapes but also the magnitudes.
The GAM\ \eqref{eq:gam} can be understood as a dynamical model inferred from the data, with which we can attempt to predict the future; it can also be understood as a histogram smoother that helps us see signal amid the noise.

\subsection{Data and code availability}The data, available at \href{https://dataverse.harvard.edu/dataset.xhtml?persistentId=doi:10.7910/DVN/B0ASZU}{Dataverse}, were analyzed using open-source software, including NumPy, pandas, SciPy, and pyGAM\ \cite{pyGAM}. Code to reproduce this work is available at \href{https://github.com/cbrummitt/machine_learned_patterns_in_economic_development}{GitHub}.


\section{Results\label{sec:results}}
\subsection{
How a machine summarizes an economy\label{sec:principal_components}}
The first principal component is the direction in product space along which countries are most spread out in terms of variance\ \cite{Lever2017}. We find this direction to be associated with a measure of diversification, as we show below.  
This first component 
explains more than half of the variation in export baskets ($57.2\%$) across countries and years. 

Mathematically, a country's score $\scoreFirstPC$ on the first principal component is a weighted sum of $\RpopSymbolCenteredScaled{c}{p}{t}$ (scaled absolute advantage) across all products. 
The weights are fixed once PCA has been fitted, but since export baskets change in time, scores $\scoreFirstPC$ change from year to year (and from country to country). 
The weights of this principal component are all positive and are depicted in the top row of Fig.\ \ref{fig:pca_loadings}(A). 
The score $\scoreFirstPC$ is highly correlated with per-capita exports summed across products, $\sum_p \exports{c}{p}{t}/\population{c}{t}$ (Pearson $\rho = 0.82$; Fig.\ SI-6). 
This correlation, however, is trivial and unsurprising given that $\scoreFirstPC$ is a positively-weighted sum across products.
However, 
the finding that the weights are all positive, together with the fact that this principal component explains almost $60\%$ of the variation, is not trivial and is of economic significance. 
It implies that in the dimension defined by the first principal component, countries are separated by their export diversification. Indeed, values of $\scoreFirstPC$ are most correlated with existing notions of diversification, once we control for other covariates which include Worldwide Governance Indicators and measures of educational attainment (see Fig.\ SI-6--SI-10).


To see why we refer to the score $\scoreFirstPC$ on the first principal component as ``complexity-weighted diversification'' (as opposed to simply ``diversification''), notice in the top row of Fig.\ \ref{fig:pca_loadings}(A) that
the loadings are not uniform: they are about twice as large on the more complex products. These variations in the loadings, in fact, are highly correlated with the Product Complexity Index\ \cite{Hidalgo2009} (Pearson $\rho = 0.81$; 
Fig.\ SI-5). 
The score $\scoreFirstPC$, in other words, captures the diversification\ \cite{Hidalgo2009} of an economy, with an emphasis on more complex goods.  
%

The next direction that most spreads out export baskets across countries and across time---conditional on being orthogonal to the first component---loads highly on machinery and negatively on agricultural products [Fig.\ \ref{fig:pca_loadings}(A), middle row]. 
Thus, after knowing a country's complexity-weighted diversity $\scoreFirstPC$, the next characteristic that most spreads out countries is how much more they export in machinery relative to agricultural goods. This second principal component explains $5\%$ of the variance, $11$ times less than that explained by $\scoreFirstPC$.

Garments have long been considered to be the first sector to industrialize in a country, 
including in England during the industrial revolution and in many East Asian countries since the 1960s\ \cite{Birdsall1993}. However, these products are not an important direction of variation of export baskets between $1962$--$\finalYearTrainingSet$ in the first and second principal components. Only in the third principal component are textile products substantially loaded [Fig.\ \ref{fig:pca_loadings}(A), bottom row]. Because this component only explains $3.3\%$ of the variance of export baskets, 
hereafter we focus on the first two principal components. 

\begin{figure*}[htb]
\begin{center}
\includegraphics{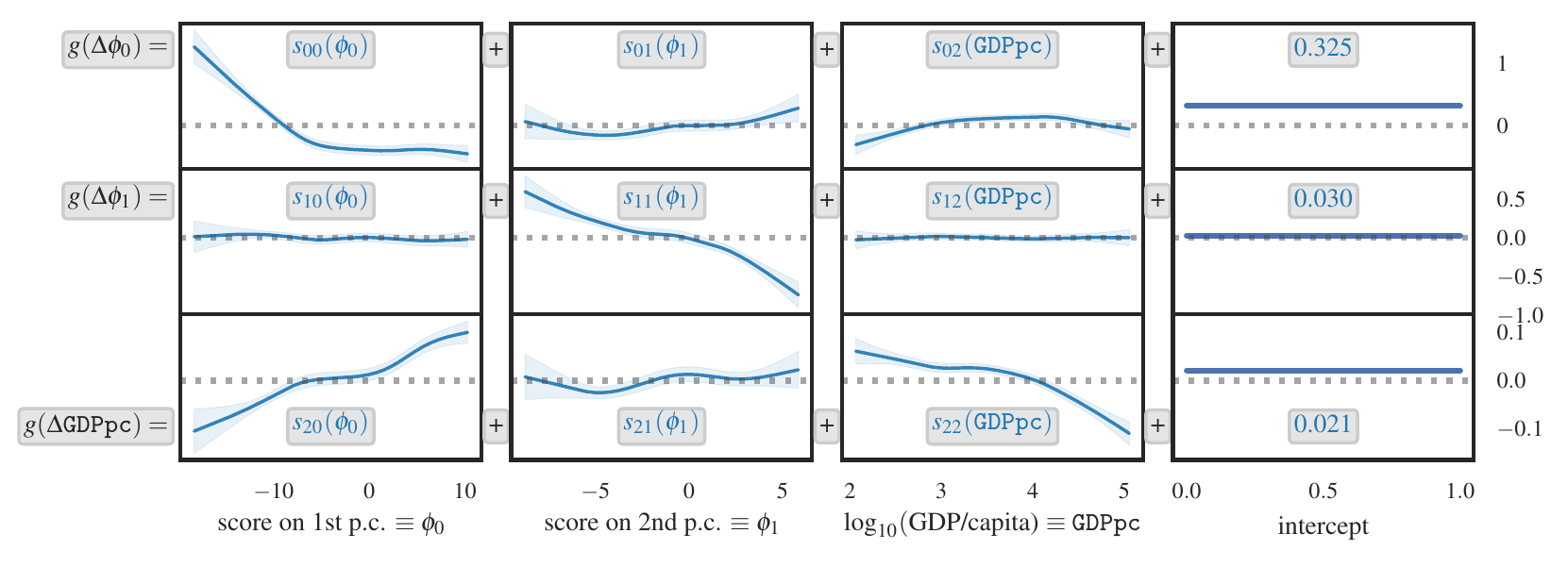}
\caption{
\textbf{
Export baskets tend to diversify and converge to a balance of agriculture and manufactured goods.
}
Shown are \emph{partial dependence plots} of the three equations in\ \eqref{eq:gam}. 
Each blue curve is an additive contribution to the quantity written in black on the left-hand side of this figure, which is a link function $g$ applied to the expected yearly change in one of the three variables $\scoreFirstPC, \scoreSecondPC, \gdppc$. 
(See the text after\ \eqref{eq:gam} for the definition of $g$.) 
In each plot, the quantity being plotted is written in blue. 
Adding the blue expressions across a row gives the right-hand sides of\ \eqref{eq:gam}. 
The shaded regions show the 95\% CI. 
Each equation has an intercept, $\gamIntercept{i}$, shown in the right column. 
The plots on the diagonal have negative trends, suggesting convergence. 
Interestingly, income is not associated with changes in export baskets, but $\scoreFirstPC$ appears to drive $\gdppc$:
diversifying precedes income growth. 
}
\label{fig:partial_dependence}
\end{center}
\end{figure*}

\subsection{Complexity-weighted diversity predicts growth\label{sec:predictors_growth}}
The partial dependence plots in Fig.\ \ref{fig:partial_dependence} show how yearly changes in $\scoreFirstPC$, $\scoreSecondPC$, and $\gdppc$ are predicted by sums of one-dimensional functions of those same variables. The rows of Fig.\ \ref{fig:partial_dependence} depict the three equations\ (\ref{eq:pc0})--(\ref{eq:gdppc}).
Notice that per-capita income is not a strong predictor of changes in export baskets as measured by $\scoreFirstPC$ and $\scoreSecondPC$ (see the top and middle plots in the third column of Fig.\ \ref{fig:partial_dependence}). 
In contrast, the score $\scoreFirstPC$ on the first principal component is associated with significant growth in income, even though $\scoreFirstPC$ and $\scoreSecondPC$ were defined independently of income. 
The fact that $\scoreFirstPC$, the complexity-weighted diversity of an economy, seems to drive income, and not the reverse, is consistent with the hypothesis that income is the outcome: income emerges from the productive capabilities of an economy, captured here by $\scoreFirstPC$ and $\scoreSecondPC$.


If the ``off-diagonal'' terms $\smoothTerm_{01} \left (\scoreSecondPC \right ) + \smoothTerm_{02} \left (\gdppc \right )$ in\ \eqref{eq:pc0} were absent, then $\scoreFirstPC$ would settle onto a value near $-10$. This amount is approximately the value of $\scoreFirstPC$ of the poorest countries in $2016$, such as Liberia, Angola, and the Democratic Republic of the Congo. However, the large intercept in\ \eqref{eq:pc0} (the top-right plot in Fig.\ \ref{fig:partial_dependence}) suggests a general positive tendency to diversify, regardless of the country's absolute advantage in machinery relative to agriculture ($\scoreSecondPC$). Once a country has complexity-weighted diversification $\scoreFirstPC > 0$, it can expect significant growth in income. 
Interestingly, simply exporting more per capita, regardless of the allocation across products, is not associated with growth: when $\scoreFirstPC$ is substituted with total per-capita exports, the relationship with income growth flattens 
(Fig.\ SI-14). 
Exporting more kinds of goods matters: Replacing $\scoreFirstPC$ with another notion of diversity \cite{Hidalgo2009} preserves the positive relationship with income growth.

We note, in addition, that the lack of clear association between $\smoothTerm_{21} \left (\scoreSecondPC \right )$ and growth of $\gdppc$ implies that there are weak returns to specialization. In fact, $\scoreSecondPC$ tends toward zero, regardless of the other two variables, meaning that countries tend toward a diversified export basket that balances agriculture with machinery. 
These results suggest that export baskets tend to increasingly resemble one another. Next we examine this convergence in more detail. 

\subsection{2D projections of the data and of the learned dynamics}

\begin{figure*}[htb]
\begin{center}
\includegraphics{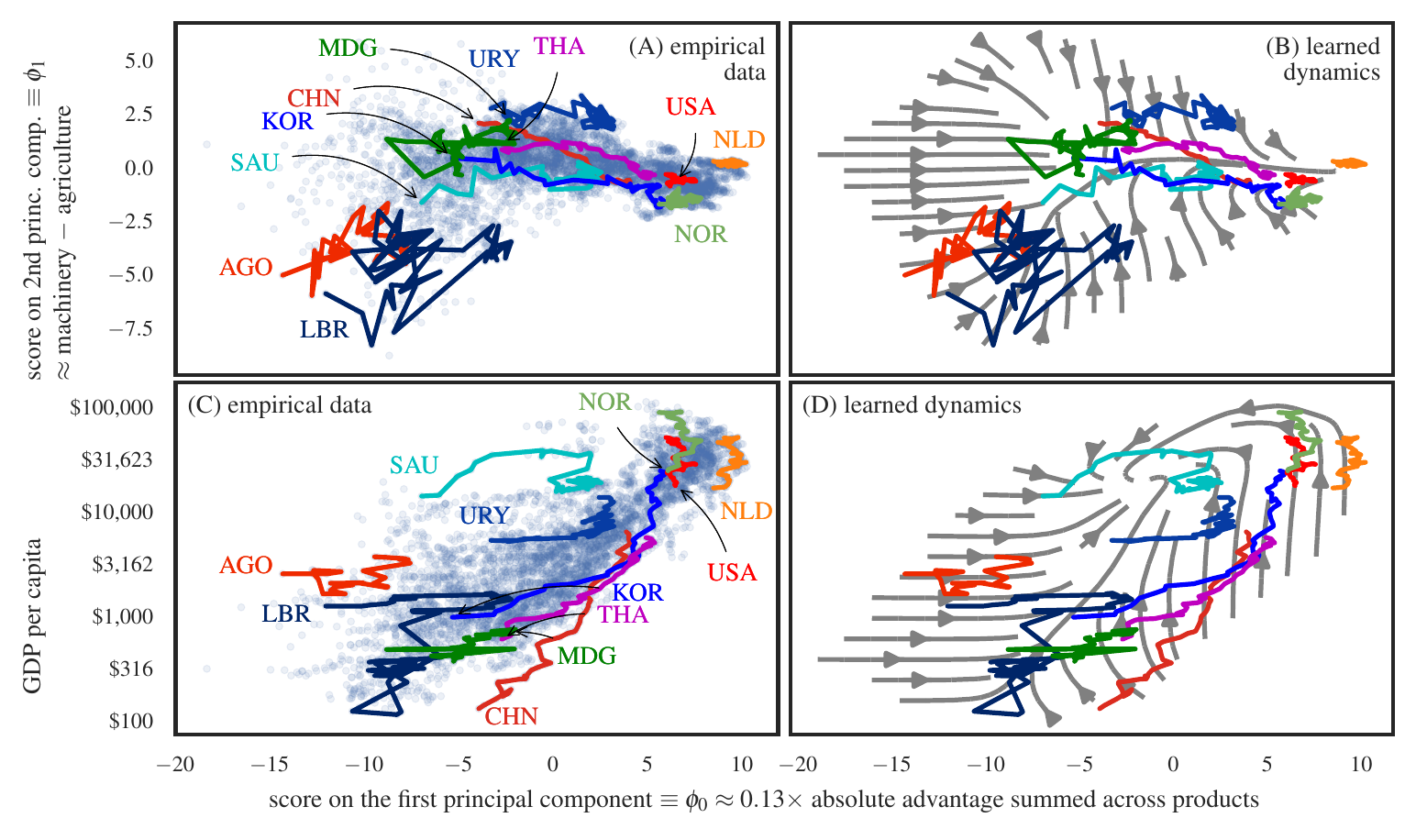}
\caption{
\textbf{The learned dynamics\ \eqref{eq:gam} 
predict that countries are converging.} 
The left column shows empirical data with blue dots; the right column shows predictions of the model\ \eqref{eq:gam} as stream plots. The empirical trajectories of eight countries over years 1962--2016 are superimposed on all four plots. 
Trajectories are labeled at the first available sample (year $1985$ for Angola, $1962$ for the rest). 
A country is represented by a triple $(\scoreFirstPC, \scoreSecondPC, \gdppc)$, and the model\ \eqref{eq:gam} has been trained on this $3$-dimensional space, but here we show projections onto $(\scoreFirstPC, \scoreSecondPC)$ in the top row and onto $(\scoreFirstPC, \gdppc)$ on the bottom row. 
(A) Countries tend to diversify (increase $\scoreFirstPC$) and strike a balance between machinery and agriculture ($\scoreSecondPC \approx 0$).  
(C) Development success stories (e.g., THA, KOR, CHN) share a common trajectory of increasing $\scoreFirstPC$ and income. 
(D) Poor countries may follow in their footsteps, but income in the richest countries may stagnate or even fall. Countries are labeled with United Nations ISO-alpha3 codes.
}
\label{fig:streamplot_scatterplot_pc0_p1_and_pc0_gdppc}
\end{center}
\end{figure*}

In Fig.\ \ref{fig:streamplot_scatterplot_pc0_p1_and_pc0_gdppc} we compare the model\ \eqref{eq:gam} with the empirical data.
This figure projects the data onto $(\scoreFirstPC, \scoreSecondPC)$ (top row) and onto $(\scoreFirstPC, \gdppc)$ (bottom row). The left-hand column shows empirical data, with some countries' trajectories highlighted. The right-hand column visualizes the vector field of\ \eqref{eq:gam} as a ``stream plot'', with the third variable not plotted taken to be the pre-$\finalYearTrainingSet$ mean. 
That is, the arrows are the expected movement for countries whose third variable (the one not plotted) equals the pre-$1988$ mean; for other countries, the arrows approximate their expected movement.

The data in Fig.\ \ref{fig:streamplot_scatterplot_pc0_p1_and_pc0_gdppc}(A) show that countries tend to move from left to right (they export more and diversify) and toward the middle of vertical axis (they move toward $\scoreSecondPC \approx 0$, a balance between agriculture and machinery). The gray streamlines of the inferred model in Fig.\ \ref{fig:streamplot_scatterplot_pc0_p1_and_pc0_gdppc}(B) confirm this pattern, suggesting that countries converge toward the trajectory like that of Thailand's (purple, labeled THA). In the bottom row, the data in Fig.\ \ref{fig:streamplot_scatterplot_pc0_p1_and_pc0_gdppc}(C) show that development success stories like South Korea (KOR), Thailand, and China (CHN) share a common trajectory of increasing $\scoreFirstPC$ and income. The inferred model's streamlines in Fig.\ \ref{fig:streamplot_scatterplot_pc0_p1_and_pc0_gdppc}(D) suggest that poor countries will follow in their footsteps, but also that income in the richest countries may fall. The ``J'' shape in Figs.\ \ref{fig:streamplot_scatterplot_pc0_p1_and_pc0_gdppc}(C) and\ \ref{fig:streamplot_scatterplot_pc0_p1_and_pc0_gdppc}(D) suggests that growth only takes over after diversification reaches a critical value. 

\subsection{Stream plots of export baskets at different levels of income}

\begin{figure*}[htb]
\begin{center}
\includegraphics{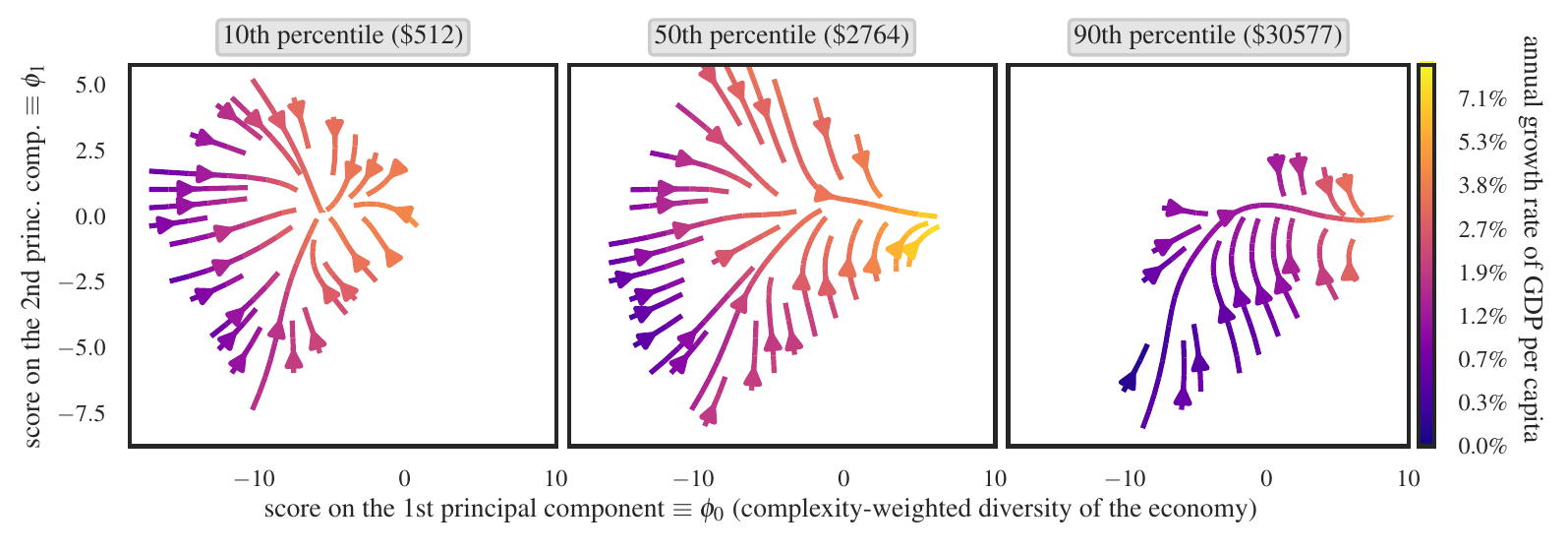}
\caption{
\textbf{Inferred dynamics of export baskets, at three levels of per-capita income, predict convergence in the long run.} The streamlines show how a country's export basket, described by its scores $(\scoreFirstPC, \scoreSecondPC)$ on the first two principal components, changes over time according to the GAM\ \eqref{eq:gam}. From left to right, the columns correspond to GDP per capita at the $10$th, $50$th, and $90$th percentiles of per-capita income among countries in the year $1988$. Those percentiles are the value inserted into\ \eqref{eq:gam}; we show streamlines at $(\scoreFirstPC, \scoreSecondPC)$ pairs in the convex hull of all empirical samples $(\scoreFirstPC, \scoreSecondPC, \gdppc)$ with $\gdppc$ within $15\%$ of the value shown at the top of the plot. The predicted yearly change in per-capita income is plotted in color. The model predicts that poor countries move toward a balance of agriculture and machinery before increasing their total exports. (Said formally, $\scoreSecondPC \to 0$ in the left plot, and $\scoreFirstPC$ increases substantially in the middle and right plots.) Eventually, all countries are predicted to become rich and to have diverse export baskets (high $\scoreFirstPC$) that balance between agriculture and machinery ($\scoreSecondPC \approx 0$).
}
\label{fig:streamplots_vary_gdppc}
\end{center}
\end{figure*}

In Fig.\ \ref{fig:streamplots_vary_gdppc} we vary $\gdppc$ across three values, the $10$th, $50$th, and $90$th percentiles of per-capita income in year $\finalYearTrainingSet$. As a country's per-capita income rises, the map of how its export basket moves through the space of products (as described by $\scoreFirstPC, \scoreSecondPC$) morphs from the plot on the left to the plot on the right. The colors denote the model's predicted change in per-capita income [\eqref{eq:gdppc}].  In the plot on the left, we see that the poorest countries tend toward a fixed point: what little they export ($\scoreFirstPC \approx -8$) tends toward a balance between agriculture and machinery ($\scoreSecondPC$ tends to zero). Countries with per-capita income near the median ($\$2764$ per year) tend to grow their complexity-weighted diversification $\scoreFirstPC$ (notice the trend to the right in the middle plot of Fig.\ \ref{fig:streamplots_vary_gdppc}), a pattern that continues for the richest countries (right-hand plot of Fig.\ \ref{fig:streamplots_vary_gdppc}). It appears that one need not be very rich to begin to diversify.

This movement in product space $(\scoreFirstPC, \scoreSecondPC)$ appears to 
maximize expected short-run increases in income, according to\ \eqref{eq:gam} 
(Fig.\ SI-18). 
High-income countries tend to be best at moving toward higher income (except for brief periods), 
and China has been exceptional at it since $1990$. 

\subsection{Long-run predictions of per-capita income}

\begin{figure}[htb]
\begin{center}
\includegraphics[width=\columnwidth]{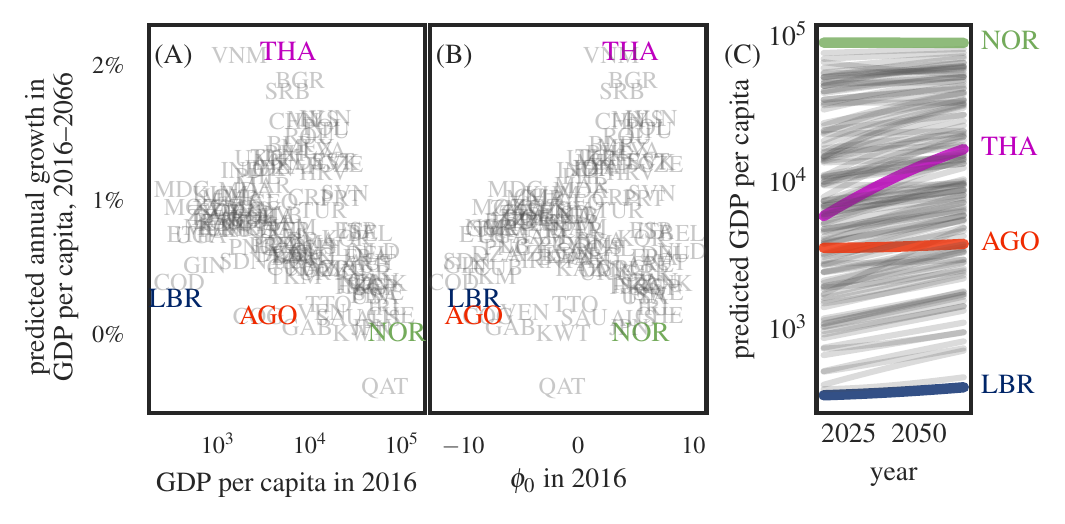}
\caption{
\textbf{Catch-up of the diverse, middle-income countries.} 
Shown 
are predicted annual growth rates of per-capita income (in constant 2010 USD per person per year) over the next $50$ years as a function of (A) current per-capita income and (B) current score $\scoreFirstPC$ on the first principal component. (C) shows predicted trajectories of per-capita income. Highlighted are four countries representative of four groups: low-income countries predicted to grow little (Liberia, LBR); middle-income countries with high diversity (high $\scoreFirstPC$) today predicted to grow a lot (Thailand, THA); middle-income countries with low diversity (low $\scoreFirstPC$) predicted to grow little (Angola, AGO); and high-income countries predicted to grow little (Norway, NOR). The GAM\ \eqref{eq:gam} predicts the highest growth in income for economies that currently have intermediate income (annual growth $\approx 1.5\%$ to $2\%$ for countries with yearly per-capita income between  $\$1000$ and $\$20{,}000$) and lower growth rates for poorest countries ($0$ to $1\%$ growth) and the richest countries ($0$ to $0.5\%$ growth).}
\label{fig:predicted_growth_rates}
\end{center}
\end{figure}

Research on economic complexity has focused on growth predictions as validation\ \cite{Hidalgo2009,Cristelli2015} and, recently, research\ \cite{Tacchella2018} has benchmarked these predictions against those of the International Monetary Fund. We found that predicting the change of export baskets simultaneously with the change of per-capita income was inherently a hard problem across different statistical learning methods [Sec.\ SI-5A.2]. Instead, we found low-dimensional models to be better suited for generating interpretable, qualitative insights rather than making competitive predictions. With this caveat in mind, we investigate the model's long-run predictions by iterating 1-year predictions starting from $2016$ data.

Figure\ \ref{fig:predicted_growth_rates} shows the model's long-run predictions of growth in per-capita income as a function of (A) per-capita income and (B) $\scoreFirstPC$ in $2016$. The model predicts that the diverse, middle-income countries today will significantly catch up to the richest ones, growing at an annual rate of $2\%$. Meanwhile, it predicts that poor countries (such as Liberia) and middle-income countries with low diversity $\scoreFirstPC$ (such as Angola) are predicted to grow between $0$ and $1\%$ annually. Rich countries like Norway are predicted to barely grow at all. The next economic success stories, according to this model, are those with intermediate income and diversification today. These results are consistent with one of the ``New Kaldor facts''\ \cite{Jones2010_NewKaldorFacts} that rich countries grow more slowly than middle income countries.

\section{Discussion\label{sec:discussion}}

This investigation sits at the intersection of three recent developments in the quantitative social, natural, and physical sciences: (1) the roles of complexity and diversity as drivers of economic growth\ \cite{Hidalgo2007,Hidalgo2009}; (2) identifying universal, low-dimensional patterns of complex human systems over time\ \cite{turchin2018quantitative}; (3) using machine learning to uncover governing laws of biological and physical systems\ \cite{Daniels2015,Zhang2015,Brunton2016_SINDy}.

We accordingly developed a new method, \method\ (\methodAcronym), by applying tools from statistical learning---namely, dimension reduction and generalized additive models---to identify stylized patterns in economic development. Our measure of 
countries' proficiencies in exporting $59$ product categories allows for small and large countries to be comparable, adjusts for global shocks, and can account for absolute economic growth. Given this data, 
\methodAcronym\ found a complexity-weighted measure of diversity, and it approximately recovered the Product Complexity Index\ \cite{Hidalgo2009}.

Our analysis generated two core insights. First, diversity appears to drive per capita income rather than the other way around. Second, countries are not predicted to split into rich and poor clubs, nor into manufacturing hubs and agricultural hubs, but instead to converge on the same increasingly diverse basket of goods (and capabilities). We hope that future research reconciles these patterns of diversification with the specialization predicted by Ricardian theories of comparative advantage\ \cite{arkolakis2012new,eaton2012putting}. The most dynamic economies of the $21^\text{st}$ century are predicted to be middle-income economies that are somewhat diversified across products.  
The least diversified countries have dismal prospects for economic growth, consistent with previous findings\ \cite{Hidalgo2007}. 

The importance of this approach rests on its applicability beyond the specific case studied here. In general, systems whose evolution is described by a multiplicity of properties are amenable to analysis such as the one we propose here. For example, it is known that wildfires typically reduce the number of species that inhabit an ecosystem, but then species recolonize over time as diversity rises in a process called ecological succession\ \cite{Rozbook}. The composition of species in a system can also converge due to migration\ \cite{Pickbook}. \methodAcronym\ could reveal other patterns, still unknown, in such ecological systems. These commonalities suggest the possibility of general theories of complex systems, uncovered by machines less tied to disciplinary paradigms.

Rapidly advancing ways for machines to learn interpretable models bode well for followup studies. Pairwise interactions could be modeled using GA$^2$M\ \cite{Lou2013_GA2M}, or high-dimensional data could be 
fitted with GAMs that both smooth the data and select terms, such as GAMSEL\ \cite{Chouldechova2015_GAMSEL} or SPLAM\ \cite{Lou2016_SPLAM}. 
 \method\ (\methodAcronym) takes a step toward this broader goal of using machines to generate fundamental theories of complex natural and social systems.




\begin{description}
 \item[Acknowledgments] C.D.B. and M.H.B. acknowledge funding from the James S. McDonnell Foundation for the Postdoctoral Award and the Scholar Award (respectively) in Complex Systems.
 \item[Competing Interests] The authors declare that they have no
competing financial interests.
 \item[Correspondence] Correspondence and requests for materials
should be addressed to C.D.B.~(email: brummitt@gmail.com).
\end{description}





\newpage

\onecolumn
\begin{center}
{\huge Supplementary Information}
\end{center}

\setcounter{figure}{0}
\setcounter{section}{0}
\makeatletter 
\renewcommand{\thefigure}{SI-\@arabic\c@figure}
\renewcommand{\theequation}{SI-\@arabic\c@equation}
\renewcommand{\thetable}{SI-\@arabic\c@table}
\renewcommand{\thesection}{SI-\@arabic\c@section}
\makeatother

\section{Related work\label{sec:related_work}}
Recent work in economics has embraced the multidimensional nature of an economy by compressing information about the products that the economy exports. The ``complexity index''\ \cite{Hidalgo2009,a2011,Albeaik2017improving} and ``fitness''\ \cite{Tacchella2012,Cristelli2013,Tacchella2013,Cristelli2013,Cristelli2015} of an economy summarize the sophistication of its capabilities. 
These measures can be defined in many ways\ \cite{Albeaik2017_729}; what they share in common is essentially a sum over products weighted by some notion of difficulty of producing that product. 
They all tackle an ambitious challenge: to describe an economy's complexity with just one number.

Other work is investigating how compressible economies and societies are. 
Machado and Mata\ \cite{Machado2015} reduce the dimensions of four time-series (per-capita income, exports relative to income, school enrollment, and lifetime expectancy) to two dimensions using multidimensional scaling. 
Hruschka et al.\ \cite{Hruschka2017} create multidimensional models of wealth by reducing the dimensions of responses to household surveys about ownership of assets such as TVs, land, and electricity. 
Turchin et al.\ \cite{turchin2018quantitative} find that societies across millennia tended to move along a common, low-dimensional trajectory in which the complexity of social organization steadily increased over time.

The existence of common patterns in the trajectories of economies can enable forecasts using simple models. 
For example, economists have fit Markov chains\ \cite{Quah1993} (and continuous versions of them\ \cite{Quah1996} and variants of Markov chains with constraints from growth theory\ \cite{Azariadis2003}) to time-series data on per-capita incomes. 
They used the stationary distribution to predict whether countries will converge to similar incomes, or whether they diverge to different ``convergence clubs''. References \cite{Quah1993,Quah1996,Azariadis2003} predict a bimodal distribution of incomes in the future. What is needed are models inferred from high-dimensional data\ \cite[p.\ 42 in Sec.\ 4.1]{Azariadis2005}.
Stochastic methods have also been used to model changes in global exports of many products\ \cite{Caraglio2016}.

Our approach was inspired by recent advances in statistical machine learning aimed at identifying governing laws of motion in data. 
One method, called SINDy\ \cite{Brunton2016_SINDy}, expands features using a hand-picked library of functions and then selects among them using sparse regression. It has since been extended to partial differential equations\ \cite{Rudy2017}, differential equations with rational terms\ \cite{Mangan2016}, information criteria\ \cite{Mangan:2017wa}, and control problems\ \cite{Brunton2016_SINDYc}. 
SINDy has proved successful in discovering laws of physics and microbiology, where we can expect polynomials and other simple functions. 
We found SINDy challenging to work well with noisy economic data with significant outliers, and great care must be taken in choosing the library of functions so that iterated predictions of the future do not diverge. 
Other approaches to system identification have used symbolic regression and genetic algorithms\ \cite{Bongard2007,Schmidt2009}; least angle regression\ \cite{Zhang2015}; and nested hierarchies of models of smooth, nonlinear dynamics\ \cite{Daniels2015}. 

Forecasting economic time-series has a long history, with the method of choice often being autoregressive--moving-average (ARMA) models\ \cite{box2015time}. 
Like the ``diffusion index'' (or ``factor augmented forecasts'')\ \cite{Bai2008}, we use principal components to reduce dimensions.

\section{Data on exports\label{sec:data}}
Our data has three main stages, which we will refer to as the \emph{raw} data, the \emph{cleaned and standardized} data, and the \emph{final, aggregated} data. 
\begin{enumerate}
\item The raw data is the data one can download freely from the United Nations' Commodities Trade Statistics website; 
\item the cleaned and standardized data is after the raw data has been expressed using standard classifications across years, and problems of the reliability of the raw records addressed and corrected; 
\item the final aggregated data is after we have removed countries and products, and then aggregated into higher level product codes, all with the goal of having reliable statistics. 
\end{enumerate}

Under a ``Premium Site License'' that Harvard has with the United Nations' Commodities Trade Statistics (COMTRADE), we provide our clean and standardized data, free to download, through the following link: 
\begin{description}
\item 
\href{http://atlas.cid.harvard.edu/}{http://atlas.cid.harvard.edu/}, specifically on \href{https://intl-atlas-downloads.s3.amazonaws.com/index.html}{this page} (\url{https://intl-atlas-downloads.s3.amazonaws.com/index.html}) by clicking on 
\href{
	https://intl-atlas-downloads.s3.amazonaws.com/CPY/S2_final_cpy_all.dta
}{S2\_final\_cpy\_all.dta} (a 451 MB file).
\end{description}

COMTRADE, the original source of our raw data, is the repository of the official trade transactions between importers and exporters. Products traded are codified in three different commodity classifications, but we express all transactions using the Standard International Trade Classification (SITC) system, Revision 2, because it covers the longest span of time. The cleaned and standardized data that we provide through the link above consists of approximately 8 million rows, each representing what a country exported of a 4-digit coded product in a year. Countries are coded following the International Organization for Standardization (ISO). We have a total of 231 unique country codes, 781 unique product codes, and 53 years (\beginningYearTrainingSet--2016).

One of the main issues with the raw data is that different countries use different classifications, and even when an importer and an exporter use the same classification, they may use different revisions.\footnote{SITC codes have had four revisions: SITC Rev.~1 in 1961, Rev.~2 in 1975, Rev.~3 in 1988, and Rev.~4 in 2006.} Each transaction in the raw data reports the code as was originally submitted by each party. Hence, to analyze the data one has to standardize the records into a single classification. COMTRADE provides concordance tables that can be used to express a product from one classification to another (\url{http://unstats.un.org/unsd/cr/registry/regdnld.asp}). But since concordance tables are typically ``many-to-many'' mappings rather than ``one-to-one'', the act of re-expressing data from one classification to another introduces additional noise because one must make some arbitrary decisions for how to split the data.

As a consequence, to get the cleaned and standardized data, the raw data has been transformed through a long process of \emph{correction} of reported transactions, \emph{cleaning} of misreported records, and \emph{standardization} of country and product codes. The process is described in detail in \cite{BustosYildirim2018}. The general approach to do this is referred to in the literature as ``mirroring'' \cite{Bhagwati1964underinvoicing,Naya1969accuracy,Yeats1978,Yeats1990,RozanskiandYeats1994,Gehlhar1996,MakhoulandOtterstrom1998,Beja2008,FerrantinoandZhi2008,Barbierietal2009,GaulierandZignago2010,Dong2010,Ferrantinoetal2012}. Mirroring consists of reconciling between what exporters and importers report, since each transaction should in principle be reported twice. But the difference between previous efforts for creating a trade dataset for research (e.g., the National Bureau of Economic Research [NBER] dataset, and the Centre d'\'{E}tudes Prospectives et d'Informations Internationales [CEPII] BACI dataset) and that of Bustos and Yildirim\ \cite{BustosYildirim2018} is that the latter accounts for transaction costs and restrictions implicit in trade reports, and they develop indices of reliability for importers, exporters, and products, which enable them to correctly impute exports of small and developing countries. Thus, the dataset of Bustos and Yildirim\ \cite{BustosYildirim2018} is more complete because it increases the number of countries with available data and additional country-product combinations (see \cite{BustosYildirim2018}), even at very disaggregated levels of the product classification.

\section{Preprocessing the exports data\label{sec:preprocessing}}
Preprocessing the exports data occurs in five steps described below:

\begin{enumerate}
\item Remove some products and countries (Sec.\ \ref{sec:filtering})
\item Normalize by population and by global exports (itself normalized by global population) (Sec.\ \ref{sec:normalize_by_population})
\item Apply a logarithmic transformation that preserves zero values and that preserves the number of values above 1 (Sec.\ \ref{sec:log_transform})
\item Center and scale for each product (Sec.\ \ref{sec:center_scale})
\item Reduce dimensions (Sec.\ \ref{sec:reduce_dimensions})
\end{enumerate}

\subsection{Filtering countries and products\label{sec:filtering}}
First, we filter the data by removing small countries and products that are not exported widely enough. The filters are similar to those in\ \cite{Albeaik2017improving} with some differences. One difference is that we avoid path dependence of the filters: we take the union of the countries and products selected by each filter, and then we remove those countries and products all at once. Another difference is that we chose not to set to zero all export values below a certain small threshold (such as $\text{US}\$5000)$ so that we do not discard information; we let the models handle noisy, small values rather than choose an arbitrary threshold. The last difference is that we remove products that have first digit in their SITC classification equal to either \productCode{3} (\emph{Fuels, lubricants \& related materials}) or \productCode{9} (\emph{Other}), which includes products such as zoo animals, coins, and gold). 

The steps below completely specify our filtering of products and countries. Countries are specified by their ISO-3166-1 alpha-3 country codes, while products are specified using the SITC classification, both found in\ \cite{TradeDataCID}.

\begin{enumerate}
\item Initialize $\countriesToRemove = \varnothing$ and $\productsToRemove = \varnothing$ (the empty set).
\item
\textbf{Remove countries with a small population}:
Select the countries with population less than 1.25 million in 2008. This selection results in the following list of $81$ countries:
\begin{quote}
$\countriesToRemove := \countriesToRemove \, \cup \, \{${ABW, AIA, AND, ANS, ANT, ASM, ATA, ATF, ATG, BHR, BHS, BLZ, BMU, BRB, BRN, BTN, BVT, CCK, COK, COM, CPV, CXR, CYM, CYP, DJI, DMA, ESH, FJI, FLK, FRO, FSM, GIB, GNQ, GRD, GRL, GUM, GUY, IOT, ISL, KIR, KNA, LCA, LUX, MAC, MDV, MHL, MLT, MNE, MNP, MSR, MUS, MYT, NCL, NFK, NIU, NRU, PCN, PLW, PYF, SGS, SHN, SLB, SMR, SPM, STP, SUR, SWZ, SYC, TCA, TKL, TLS, TON, TUV, TWN, UMI, VAT, VCT, VGB, VUT, WLF, WSM}$\}$
\end{quote}
\item 
\textbf{Remove countries with little total export value}:
Select the countries with total export value smaller than $1$ billion USD in 2008. This selection results in the following $81$ countries: 
\begin{quote}
$\countriesToRemove := \countriesToRemove \, \cup \, \{${AFG, AIA, AND, ARM, ASM, ATA, ATF, ATG, BDI, BEN, BFA, BLZ, BRB, BTN, BVT, CAF, CCK, COK, COM, CPV, CXR, CYM, DJI, DMA, ERI, ESH, FJI, FLK, FRO, FSM, GIB, GMB, GNB, GRD, GRL, GUM, GUY, HTI, IOT, KIR, KNA, LCA, LSO, MDV, MNE, MNP, MSR, MWI, MYT, NER, NFK, NIU, NPL, NRU, PCN, PLW, PSE, PYF, RWA, SGS, SHN, SLB, SLE, SMR, SOM, SPM, STP, SYC, TCA, TGO, TKL, TLS, TON, TUV, UMI, VAT, VCT, VGB, VUT, WLF, WSM}$\}$
\end{quote}
\item \textbf{Remove countries that export very few products}: Select countries with zero export value for at least $95\%$ of products in some year. This selection results in the following list of $52$ countries:
\begin{quote}
$\countriesToRemove := \countriesToRemove \, \cup \, \{${AIA, ATA, ATF, BDI, BTN, BVT, CCK, COK, COM, CPV, CXR, ERI, ESH, FLK, FSM, GNB, GNQ, GUF, HMD, IOT, KIR, LAO, LCA, MDV, MHL, MNG, MNP, MRT, MTQ, NFK, NIU, NPL, NRU, PCI, PCN, PYF, RWA, SGS, SSD, STP, SYC, TCA, TLS, TON, TUV, UMI, VGB, VIR, VUT, WLF, WSM, YEM}$\}$
\end{quote}
\item \textbf{Remove war-torn countries}: Add Afghanistan (AFG), Iraq (IRQ), and Chad (TCD) to the set of countries to remove:
\begin{quote}
$\countriesToRemove := \countriesToRemove \, \cup \, \{${AFG, IRQ, TCD}$\}$
\end{quote}
\item \textbf{Remove all products in the categories of fossil fuels and miscellaneous}: Add to the set of products to remove all the products with first digit (in the SITC classification scheme) equal to \productCode{3} (fossil fuels) or \productCode{9} (miscellaneous products such as art and coins):
\begin{quote}
$\productsToRemove := \productsToRemove \, \cup \, \{{\productCode{3}^*, \productCode{9}^*}\}$
\end{quote}
Here, $\productCode{3}^*$ means any product code that begins with $\productCode{3}$.
\item \textbf{Remove products exported by few countries}: Select products not exported by at least $80\%$ of countries in at least one year. This selection results in the following $78$ product codes:
\begin{quote}
$\productsToRemove := \productsToRemove \, \cup \, \{${
\productCode{0019}, \productCode{0115}, \productCode{0451}, \productCode{0452}, \productCode{0742}, \productCode{2114}, \productCode{2223}, \productCode{2226}, \productCode{2231}, \productCode{2232}, \productCode{2234}, \productCode{2235}, \productCode{2512}, \productCode{2516}, \productCode{2518}, \productCode{2613}, \productCode{2634}, \productCode{2652}, \productCode{2654}, \productCode{2655}, \productCode{2659}, \productCode{2685}, \productCode{2712}, \productCode{2714}, \productCode{2741}, \productCode{2742}, \productCode{2784}, \productCode{2814}, \productCode{2816}, \productCode{2860}, \productCode{2872}, \productCode{2876}, \productCode{3223}, \productCode{3224}, \productCode{3231}, \productCode{3341}, \productCode{3342}, \productCode{3343}, \productCode{3344}, \productCode{3415}, \productCode{3510}, \productCode{4233}, \productCode{4236}, \productCode{4241}, \productCode{4244}, \productCode{4245}, \productCode{5163}, \productCode{5223}, \productCode{5249}, \productCode{5323}, \productCode{5828}, \productCode{6112}, \productCode{6113}, \productCode{6121}, \productCode{6344}, \productCode{6546}, \productCode{6642}, \productCode{6674}, \productCode{6727}, \productCode{6741}, \productCode{6750}, \productCode{6784}, \productCode{6793}, \productCode{6831}, \productCode{6880}, \productCode{7187}, \productCode{7433}, \productCode{7521}, \productCode{7524}, \productCode{7911}, \productCode{7912}, \productCode{7913}, \productCode{7914}, \productCode{7924}, \productCode{7931}, \productCode{8821}, \productCode{8941}, \productCode{9110}
}$\}$
\end{quote}
\item \textbf{Remove products with little global exports}: Select products with global exports $< 10$ million in some year. This selection results in the following $37$ products: 
\begin{quote}
$\productsToRemove := \productsToRemove \, \cup \, \{${
\productCode{0019}, \productCode{0742}, \productCode{1122}, \productCode{2114}, \productCode{2232}, \productCode{2235}, \productCode{2239}, \productCode{2634}, \productCode{2652}, \productCode{2711}, \productCode{2714}, \productCode{3224}, \productCode{3415}, \productCode{4311}, \productCode{5223}, \productCode{5323}, \productCode{5828}, \productCode{6112}, \productCode{6113}, \productCode{6121}, \productCode{6122}, \productCode{6349}, \productCode{6546}, \productCode{6642}, \productCode{6646}, \productCode{6674}, \productCode{6741}, \productCode{6750}, \productCode{6880}, \productCode{6912}, \productCode{7187}, \productCode{7213}, \productCode{7433}, \productCode{7521}, \productCode{7524}, \productCode{8941}, \productCode{9110}
}$\}$
\end{quote}
\item \textbf{Remove products with little market share}:
Select products whose market share is below the fifth percentile in year 2008. This selection results in the following $39$ products: 
\begin{quote}
$\productsToRemove := \productsToRemove \, \cup \, \{${
\productCode{0129}, \productCode{0742}, \productCode{2114}, \productCode{2231}, \productCode{2232}, \productCode{2235}, \productCode{2440}, \productCode{2614}, \productCode{2632}, \productCode{2640}, \productCode{2652}, \productCode{2654}, \productCode{2655}, \productCode{2659}, \productCode{2685}, \productCode{2686}, \productCode{2687}, \productCode{2712}, \productCode{2714}, \productCode{2742}, \productCode{2923}, \productCode{3231}, \productCode{3415}, \productCode{4233}, \productCode{4314}, \productCode{6112}, \productCode{6121}, \productCode{6518}, \productCode{6545}, \productCode{6576}, \productCode{6593}, \productCode{6642}, \productCode{6880}, \productCode{6932}, \productCode{7163}, \productCode{7511}, \productCode{7521}, \productCode{7612}, \productCode{7631}
}$\}$
\end{quote}
\end{enumerate}

In the end, these filters remove 121 products (listed in Tables\ \ref{tab:removed_products_1}, \ref{tab:removed_products_2}, and \ref{tab:removed_products_3}) and the following 112 countries:
\begin{quote}
Afghanistan (AFG); American Samoa (ASM); Andorra (AND); Anguilla (AIA); Antarctica (ATA); Antigua and Barbuda (ATG); Armenia (ARM); Aruba (ABW); Bahamas (BHS); Bahrain (BHR); Barbados (BRB); Belize (BLZ); Benin (BEN); Bermuda (BMU); Bhutan (BTN); Bouvet Island (BVT); British Indian Ocean Territory (IOT); British Virgin Islands (VGB); Brunei (BRN); Burkina Faso (BFA); Burundi (BDI); Cape Verde (CPV); Cayman Islands (CYM); Central African Republic (CAF); Chad (TCD); Christmas Island (CXR); Cocos (Keeling) Islands (CCK); Comoros (COM); Cook Islands (COK); Cyprus (CYP); Djibouti (DJI); Dominica (DMA); Equatorial Guinea (GNQ); Eritrea (ERI); Falkland Islands (FLK); Faroe Islands (FRO); Fiji (FJI); French Guiana (GUF); French Polynesia (PYF); French South Antarctic Territory (ATF); Gambia (GMB); Gibraltar (GIB); Greenland (GRL); Grenada (GRD); Guam (GUM); Guinea-Bissau (GNB); Guyana (GUY); Haiti (HTI); Heard Island and McDonald Islands (HMD); Holy See (Vatican City) (VAT); Iceland (ISL); Iraq (IRQ); Kiribati (KIR); Laos (LAO); Lesotho (LSO); Luxembourg (LUX); Macau (MAC); Malawi (MWI); Maldives (MDV); Malta (MLT); Marshall Islands (MHL); Martinique (MTQ); Mauritania (MRT); Mauritius (MUS); Mayotte (MYT); Micronesia (FSM); Mongolia (MNG); Montenegro (MNE); Montserrat (MSR); Nauru (NRU); Nepal (NPL); Netherlands Antilles (ANT); New Caledonia (NCL); Niger (NER); Niue (NIU); Norfolk Island (NFK); Northern Mariana Islands (MNP); Pacific Island (US) (PCI); Palau (PLW); Palestine (PSE); Pitcairn Islands (PCN); Rwanda (RWA); Saint Helena (SHN); Saint Kitts and Nevis (KNA); Saint Lucia (LCA); Saint Pierre and Miquelon (SPM); Saint Vincent and the Grenadines (VCT); Samoa (WSM); San Marino (SMR); Sao Tome and Principe (STP); Seychelles (SYC); Sierra Leone (SLE); Solomon Islands (SLB); Somalia (SOM); South Georgia South Sandwich Islands (SGS); South Sudan (SSD); Suriname (SUR); Swaziland (SWZ); Taiwan (TWN); Timor-Leste (TLS); Togo (TGO); Tokelau (TKL); Tonga (TON); Turks and Caicos Islands (TCA); Tuvalu (TUV); United States Minor Outlying Islands (UMI); Vanuatu (VUT); Virgin Islands (VIR); Wallis and Futuna (WLF); Western Sahara (ESH); Yemen (YEM).
\end{quote}

\begin{table}[htp]
\caption{121 removed products (part 1)}
\begin{center}
\begin{tabular}{c|c}
\productCode{0019} &    Live animals of a kind mainly used for human food, nes \\
\productCode{0115} &    Meat of horses, asses, mules and hinnies, fresh, chilled or frozen \\
\productCode{0129} &    Meat and edible meat offal, nes, in brine, dried, salted or smoked \\
\productCode{0451} &    Rye, unmilled \\
\productCode{0452} &    Oats, unmilled \\
\productCode{0742} &    Mate \\
\productCode{1122} &    Other fermented beverages, nes (cider, perry, mead, etc) \\
\productCode{2114} &    Goat and kid skins, raw, whether or not split \\
\productCode{2223} &    Cotton seeds \\
\productCode{2226} &    Rape and colza seeds \\
\productCode{2231} &    Copra \\
\productCode{2232} &    Palm nuts and kernels \\
\productCode{2234} &    Linseed \\
\productCode{2235} &    Castor oil seeds \\
\productCode{2239} &    Flour or meals of oil seeds or oleaginous fruit, non-defatted \\
\productCode{2440} &    Cork, natural, raw and waste \\
\productCode{2512} &    Mechanical wood pulp \\
\productCode{2516} &    Chemical wood pulp, dissolving grades \\
\productCode{2518} &    Chemical wood pulp, sulphite \\
\productCode{2613} &    Raw silk (not thrown) \\
\productCode{2614} &    Silk worm cocoons and silk waste \\
\productCode{2632} &    Cotton linters \\
\productCode{2634} &    Cotton, carded or combed \\
\productCode{2640} &    Jute, other textile bast fibres, nes, raw, processed but not spun \\
\productCode{2652} &    True hemp, raw or processed but not spun, its tow and waste \\
\productCode{2654} &    Sisal, agave fibres, raw or processed but not spun, and waste \\
\productCode{2655} &    Manila hemp, raw or processed but not spun, its tow and waste \\
\productCode{2659} &    Vegetable textile fibres, nes, and waste \\
\productCode{2685} &    Horsehair and other coarse animal hair, not carded or combed \\
\productCode{2686} &    Waste of sheep's or lambs' wool, or of other animal hair, nes \\
\productCode{2687} &    Sheep's or lambs' wool, or of other animal hair, carded or combed \\
\productCode{2711} &    Animal or vegetable fertilizer, crude \\
\productCode{2712} &    Natural sodium nitrate \\
\productCode{2714} &    Potassium salts, natural, crude \\
\productCode{2741} &    Sulphur (other than sublimed, precipitated or colloidal) \\
\productCode{2742} &    Iron pyrites, unroasted \\
\productCode{2784} &    Asbestos \\
\productCode{2814} &    Roasted iron pyrites \\
\productCode{2816} &    Iron ore agglomerates \\
\productCode{2860} &    Ores and concentrates of uranium and thorium \\
\productCode{2872} &    Nickel ores and concentrates; nickel mattes, etc \\
\end{tabular}
\end{center}
\label{tab:removed_products_1}
\end{table}%

\begin{table}[htp]
\caption{121 removed products (part 2)}
\begin{center}
\begin{tabular}{c|c}
\productCode{2876} &    Tin ores and concentrates \\
\productCode{2923} &    Vegetable plaiting materials \\
\productCode{3221} &    Anthracite, not agglomerated \\
\productCode{3222} &    Other coal, not agglomerated \\
\productCode{3223} &    Lignite, not agglomerated \\
\productCode{3224} &    Peat, not agglomerated \\
\productCode{3231} &    Briquettes, ovoids, from coal, lignite or peat \\
\productCode{3232} &    Coke and semi-coke of coal, of lignite or peat; retort carbon \\
\productCode{3330} &    Crude petroleum and oils obtained from bituminous materials \\
\productCode{3341} &    Gasoline and other light oils \\
\productCode{3342} &    Kerosene and other medium oils \\
\productCode{3343} &    Gas oils \\
\productCode{3344} &    Fuel oils, nes \\
\productCode{3345} &    Lubricating petroleum oils, and preparations, nes \\
\productCode{3351} &    Petroleum jelly and mineral waxes \\
\productCode{3352} &    Mineral tars and products \\
\productCode{3353} &    Mineral tar pitch, pitch coke \\
\productCode{3354} &    Petroleum bitumen, petroleum coke and bituminous mixtures, nes \\
\productCode{3413} &    Petroleum gases and other gaseous hydrocarbons, nes, liquefied \\
\productCode{3414} &    Petroleum gases, nes, in gaseous state \\
\productCode{3415} &    Coal gas, water gas and similar gases \\
\productCode{3510} &    Electric current \\
\productCode{4233} &    Cotton seed oil \\
\productCode{4236} &    Sunflower seed oil \\
\productCode{4241} &    Linseed oil \\
\productCode{4244} &    Palm kernel oil \\
\productCode{4245} &    Castor oil \\
\productCode{4311} &    Processed animal and vegetable oils \\
\productCode{4314} &    Waxes of animal or vegetable origin \\
\productCode{5163} &    Inorganic esters, their salts and derivatives \\
\productCode{5223} &    Halogen and sulphur compounds of non-metals \\
\productCode{5249} &    Other radio-active and associated materials \\
\productCode{5323} &    Synthetic tanning substances; tanning preparations \\
\productCode{5828} &    Ion exchangers of the condensation, polycondensation etc \\
\productCode{6112} &    Composition leather, in slabs, sheets or rolls \\
\productCode{6113} &    Calf leather \\
\productCode{6121} &    Articles of leather use in machinery or mechanical appliances, etc \\
\productCode{6122} &    Saddlery and harness, of any material, for any kind of animal \\
\productCode{6344} &    Wood-based panels, nes \\
\productCode{6349} &    Wood, simply shaped, nes \\
\end{tabular}
\end{center}
\label{tab:removed_products_2}
\end{table}%

\begin{table}[htp]
\caption{121 removed products (part 3)}
\begin{center}
\begin{tabular}{c|c}
\productCode{6518} &    Yarn of regenerated fibres, put up for retail sale \\
\productCode{6545} &    Fabrics, woven of jute or other textile bast fibres of heading 2640 \\
\productCode{6546} &    Fabrics of glass fibre (including narrow, pile fabrics, lace, etc) \\
\productCode{6576} &    Hat shapes, hat-forms, hat bodies and hoods \\
\productCode{6593} &    Kelem, Schumacks and Karamanie rugs and the like \\
\productCode{6642} &    Optical glass and elements of optical glass (unworked) \\
\productCode{6646} &    Bricks, tiles, etc of pressed or moulded glass, used in building \\
\productCode{6674} &    Synthetic or reconstructed precious or semi-precious stones \\
\productCode{6727} &    Iron or steel coils for re-rolling \\
\productCode{6741} &    Universal plates of iron or steel \\
\productCode{6750} &    Hoop and strip of iron or steel, hot-rolled or cold-rolled \\
\productCode{6784} &    High-pressure hydro-electric conduit of steel \\
\productCode{6793} &    Steel and iron forging and stampings, in the rough state \\
\productCode{6831} &    Nickel and nickel alloys, unwrought \\
\productCode{6880} &    Uranium depleted in U235, thorium, and alloys, nes; waste and scrap \\
\productCode{6912} &    Structures and parts of, of aluminium; plates, rods, and the like \\
\productCode{6932} &    Barbed iron or steel wire: fencing wire \\
\productCode{7163} &    Rotary converters \\
\productCode{7187} &    Nuclear reactors, and parts thereof, nes \\
\productCode{7213} &    Dairy machinery, nes (including milking machines), and parts nes \\
\productCode{7433} &    Free-piston generators for gas turbines and parts thereof, nes \\
\productCode{7511} &    Typewriters; cheque-writing machines \\
\productCode{7521} &    Analogue and hybrid data processing machines \\
\productCode{7524} &    Digital central storage units, separately consigned \\
\productCode{7612} &    Television receivers, monochrome \\
\productCode{7631} &    Gramophones and record players, electric \\
\productCode{7911} &    Rail locomotives, electric \\
\productCode{7912} &    Other rail locomotives; tenders \\
\productCode{7913} &    Mechanically propelled railway, tramway, trolleys, etc \\
\productCode{7914} &    Railway, tramway passenger coaches, etc, not mechanically propelled \\
\productCode{7924} &    Aircraft of an unladen weight exceeding 15000 kg \\
\productCode{7931} &    Warships \\
\productCode{8821} &    Chemical products and flashlight materials for use in photografy \\
\productCode{8941} &    Baby carriages and parts thereof, nes \\
\productCode{9110} &    Postal packages not classified according to kind \\
\productCode{9310} &    Special transactions, commodity not classified according to class \\
\productCode{9410} &    Animals, live, nes, (including zoo animals, pets, insects, etc) \\
\productCode{9510} &    Armoured fighting vehicles, war firearms, ammunition, parts, nes \\
\productCode{9610} &    Coin (other than gold coin), not being legal tender \\
\productCode{9710} &    Gold, non-monetary (excluding gold ores and concentrates)

\end{tabular}
\end{center}
\label{tab:removed_products_3}
\end{table}%

\paragraph{Final dataset}

After removing countries and products, we have a dataset of $138$ countries, $665$ products at the 4-digit level, and $6377$ distinct (country, year) pairs. Summing export values at the $2$-digit level results in $\numProducts$ products. Merging this exports data with population data from the World Bank\ \cite{GDPpcWorldBank} and from\ \cite{TradeDataCID} drops 240 (country, year) samples, resulting in $6137$ distinct (country, year) pairs. 

\begin{figure}[htbp]
\begin{center}
\includegraphics{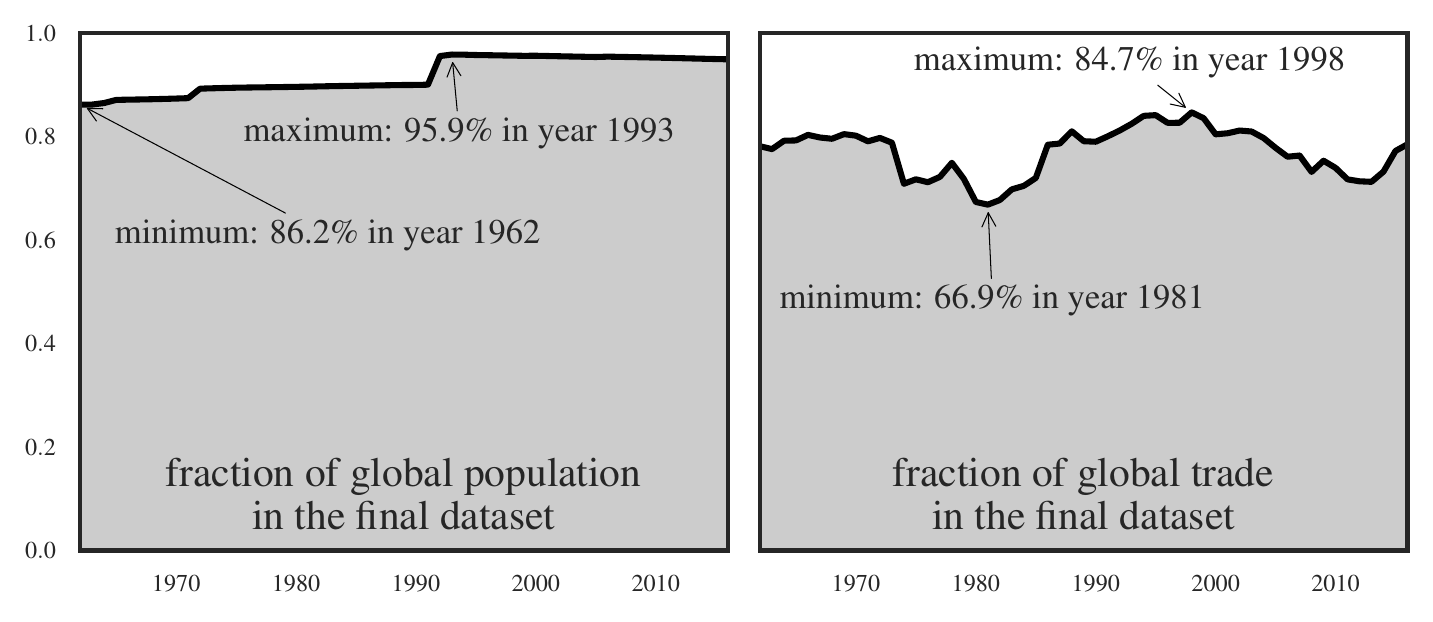}
\caption{Fraction of global population and global trade in the dataset after the filters described in Sec.\ \ref{sec:filtering} are applied.}
\label{fig:frac_global_population_trade}
\end{center}
\end{figure}

This dataset has on average $92\%$ of the global population (minimum $86\%$, maximum $96\%$) and $77\%$ of global trade (minimum $67\%$, maximum $85\%$). 
Time-series of those values are plotted in Figure\ \ref{fig:frac_global_population_trade}.

\subsection{Normalize export values by population and by global exports\label{sec:normalize_by_population}}

To make small and large countries comparable, we divide the value 
of a 
country $c$'s 
exports of a product 
$p$ in year $t$, denoted $\exports{c}{p}{t}$,
by a null model of a country's expected value of its exports of that product given that country's population, $\E \left [ \exports{c}{p}{t} \mid \population{c}{t} \right ]$. 
To remove the effects of global price shocks, we divide this quantity by the total value of the world's exports of that product, which we also normalize by a null model that predicts global export value using global population. 
Formally, for each country $c$ in a set of $123$ countries $\countries$ and for each product $p$ in the set of $\numProducts$ products $\products$, we define the \emph{absolute advantage} of country $c$ in product $p$ as 
\begin{align}
\RpopSymbol{c}{p}{t} :=
\frac
{
	\exports{c}{p}{t} / \E \left [ \exports{c}{p}{t} \vert \population{c}{t} \right ]
}
{
	\sum_c \exports{c}{p}{t} / \E \left [\sum_c  \exports{c}{p}{t} \big \vert \sum_c  \population{c}{t} \right ]
}
\label{eq:define_Rpop}
\end{align}

\subsubsection{Null models of export values based on population size}\label{sec:null_model}

Countries with more people tend to export more, but typically not in proportion to their population size. 
To allow for product-specific variation in the relationship between exports and population, we assume that the expectations in\ \eqref{eq:define_Rpop} follow power laws of population size.


A country with population double that of another country typically exports more, but rarely does it export twice as much. For intuition, consider a disk-shaped country with its population distributed evenly across space and with exports occurring at the border in proportion to the size of the perimeter. That country's exports increase with the square root of the population size. (This example is more extreme than reality: the exponent is $\approx 0.88$ rather than $0.5$.) Motivated by this intuition, we create a null model of exports by assuming that export value of a certain product, either by a certain country or by the whole world, grows with population size raised to some power, and that this exponent varies from one product to another. Specifically, we assume that
\begin{align}
\nullExportsSymbol{c}{p}{t} &= \nullExports{p},
\label{eq:power_law_numerator} \\
\nullGlobalExportsSymbol{p}{t} &= \nullGlobalExports{p}
\label{eq:power_law_denominator}
\end{align}
With\ \eqref{eq:power_law_numerator} and\ \eqref{eq:power_law_denominator}, our measure of a country $c$'s \emph{absolute advantage} 
in producing the product $p$ in year $t$ is
\begin{align}
\RpopSymbol{c}{p}{t} 
&= \RpopFraction{c}{p}{t}.
\label{eq:Rpop_written_out}
\end{align}
This quantity $\RpopSymbol{c}{p}{t}$ captures how proficient a country $c$ is in exporting product $p$ in year $t$, relative to an average country of its population size. 

\begin{figure}[htbp]
\begin{center}
\includegraphics{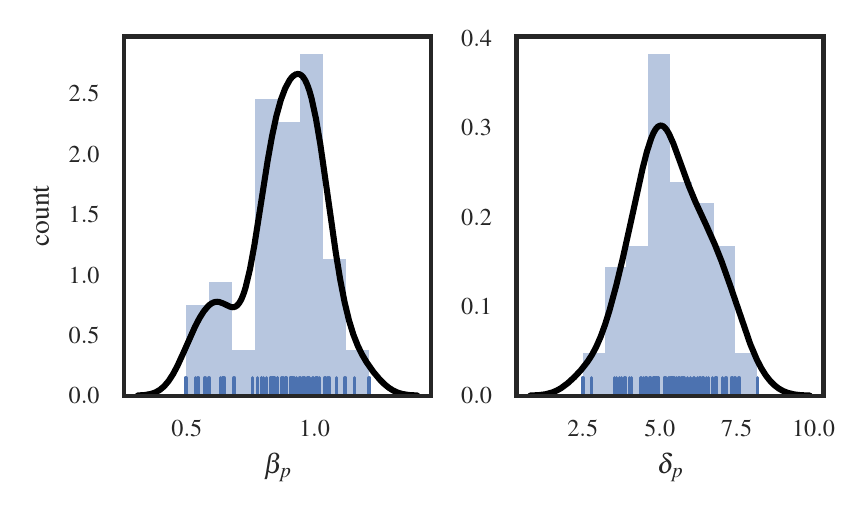}
\caption{Distribution of exponents $\countrySpecificExponent{p}$ and $\globalExponent{p}$ in the null models\ \eqref{eq:power_law_numerator} and\ \eqref{eq:power_law_denominator}, respectively. For illustrative purposes, we draw in black a kernel density estimate with a Gaussian kernel.}
\label{fig:exponents_from_normalizing_Rpop_by_population}
\end{center}
\end{figure}

The distributions of the exponents $\countrySpecificExponent{p}$ and $\globalExponent{p}$ are plotted in Figure\ \ref{fig:exponents_from_normalizing_Rpop_by_population}.
The exponents $\countrySpecificExponent{p}$ have average value of $0.88$; the minimum is $0.49$ for the product \emph{Dairy products and birds' eggs} (product code \productCode{02}), and the maximum is $1.21$ for the product \emph{Crude rubber (including synthetic and reclaimed)} (product code \productCode{23}). 
Thus, the export value of certain product tends to grow sublinearly with the population size, in accordance with the hypothetical disk-shaped country described above.
Meanwhile, the exponents $\globalExponent{p}$ are much larger: the average (across all $\numProducts$ products) is $5.37$. The minimum is $2.50$ for the product \emph{Textile fibers (not wool tops) and their wastes (not in yarn)} (product code \productCode{26}), and the maximum is $8.18$ for the product \emph{Office machines and automatic data processing equipment)} (product code \productCode{75}). 
Thus, global exports of a product tend to grow superlinearly with global population.

\subsection{Logarithmically transforming data with lots of zeros in it\label{sec:log_transform}}

\begin{figure}[htbp]
\begin{center}
\includegraphics{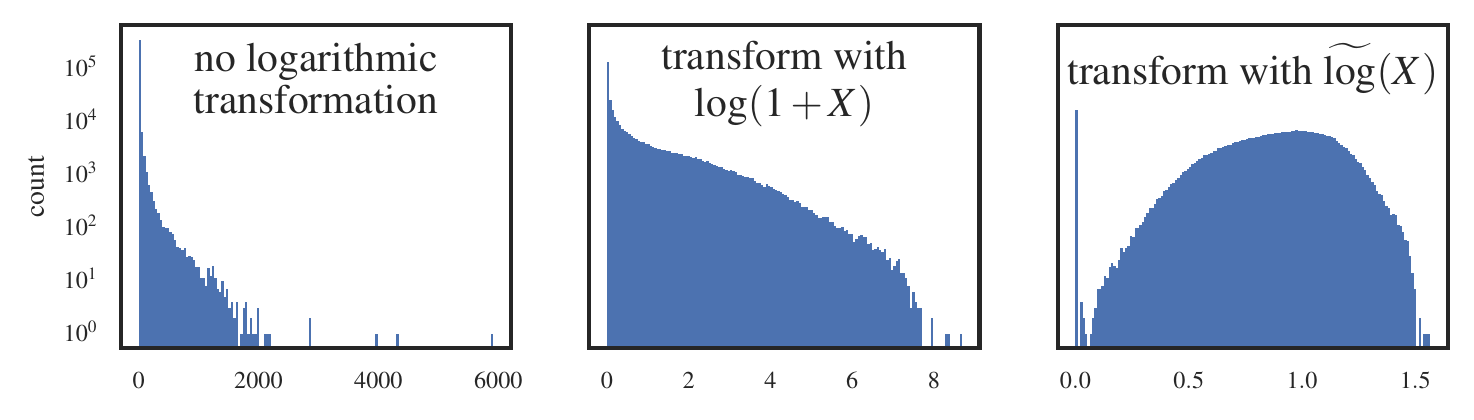}
\caption{Histograms of the flattened data\ \eqref{eq:Rpop_written_out} before it is logarithmically transformed (left panel), after it is logarithmically transformed with $\log(1 + \cdot)$ (middle plot), and after it is logarithmically transformed with $\widetilde \log(\cdot)$ (right plot).}
\label{fig:histograms_log_transformations}
\end{center}
\end{figure}

In this paper, we consider yearly export values $\exports{c}{p}{t}$ of $\numProducts$ two-digit products. These export values range from zero to nearly a trillion US dollars per year. 
China, for example, has recently exported over $\$300$ billion in \emph{Electric machinery, apparatus and appliances, nes, and parts, nes} (product code \productCode{77}) in one year.
After normalizing by population and by global exports with\ \eqref{eq:define_Rpop}, the values are still rather heavy-tailed and range from $0$ to $7.3 \times 10^4$ US dollars per year; see the left and middle panels of Figure\ \ref{fig:histograms_log_transformations}.

\begin{figure}[htbp]
\begin{center}
\includegraphics{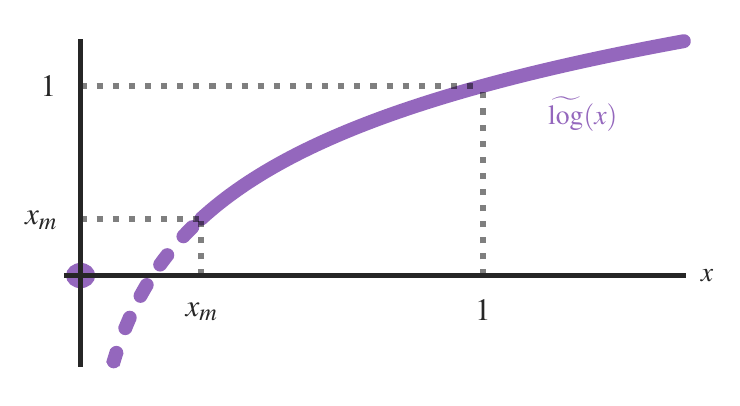}
\caption{The logarithmic transformation\ \eqref{eq:log_transformation} used. It leaves unchanged zeros and whether values are above one.}
\label{fig:plot_of_log_transformation_function}
\end{center}
\end{figure}

One way to logarithmically transform heavy-tailed data with zeros in it is to add one before applying the natural logarithm, so that zero maps to zero. However, we found that this transformation resulted in data that was approximately exponentially distributed rather than normally distributed, and we found that adding one introduces a scale in the data. To avoid these outcomes, we applied a different logarithmic transformation that is plotted in Fig.\ \ref{fig:plot_of_log_transformation_function}:
\begin{align}
\widetilde \log(x) \equiv 
\begin{cases}
1 + s \log(x) & \text{if } x > 0 \\
0 & \text{if } x = 0.
\end{cases} \label{eq:log_transformation}
\end{align}
where the scaling factor
\begin{align}
s \equiv \lim_{z \to \Xminpos} \frac { z - 1 } {\log(z) } = 
\begin{cases}
	1
		& \text{if } \Xminpos = 1 \\
	\left ( \Xminpos - 1 \right ) /  \log \Xminpos
		& \text{if } \Xminpos \neq 1
\end{cases}, \label{eq:scaling_factor_in_log}
\end{align}
and  $\Xminpos$ is the smallest positive value of all elements of the matrix $X$:
\begin{align}
\Xminpos \equiv \min \left \{x : x \in X, x > 0 \right \}.
\end{align}

The limit in\ \eqref{eq:scaling_factor_in_log} ensures that $s$ exists for all $\Xminpos > 0$; in particular, $s = 1$ when $\Xminpos = 1$.
Note that 
\begin{align}
\widetilde \log(\Xminpos) &= \Xminpos, \label{eq:equal_at_Xminpos} \\
\widetilde \log(1) &= 1, \label{eq:equal_at_one} \\
\widetilde \log(x) &\text{ is increasing}.  \label{eq:logtilde_increasing}
\end{align}
Equations\ \eqref{eq:equal_at_Xminpos} and\ \eqref{eq:equal_at_one} are direct computations. Equation\ \eqref{eq:logtilde_increasing} holds because $\left ( z - 1 \right ) / \log(z)$ is positive for $z > 0$. 
A consequence of\ \eqref{eq:equal_at_one} and of\ \eqref{eq:logtilde_increasing} is that 
\begin{align}
\widetilde \log (x) > 1 \text{ if and only if } x > 1. \label{eq:preserve_above_1}
\end{align}
Statement\ \eqref{eq:preserve_above_1} is an important property for a logarithmic transformation of data like that studied here: 
because the data is normalized by dividing by the prediction of a null model, being above one (or not) is meaningful, so we wish our logarithmic transformation to preserve which values are above one and which values are below one.

\subsection{Centering and scaling\label{sec:center_scale}}

Next we pivot the data so that the rows are 
observations of a certain country in a certain year, 
and the columns are the values of $\RpopSymbol{c}{p}{t}$ for each of the $\numProducts$ many products $p$. 
We center and scale the columns using the pre-$1989$ column means and standard deviations:
\begin{align}
\RpopSymbolCenteredScaled{c}{p}{t} :=
\frac
{
	\RpopSymbol{c}{p}{t}
	-
	\average
		\left( 
			\{\RpopSymbol{c}{p}{t} : \beginningYearTrainingSet \leq t \leq \finalYearTrainingSet, c \in \countries \} 
		\right )
}
{
	\stddev
		\left(
			\{\RpopSymbol{c}{p}{t} : \beginningYearTrainingSet \leq t \leq \finalYearTrainingSet, c \in \countries \}
		\right )
}
\label{eq:center_scale}
\end{align}
where $\average$ denotes mean and $\stddev$ denotes standard deviation. 
The column means and standard deviations, like all other preprocessing steps such as dimension reduction described next, are fit to data from year $\finalYearTrainingSet$ or earlier. That way, we can split the data into cross validation sets that are nested in time, and all preprocessing is done with the earliest set of data (years $\beginningYearTrainingSet$ to $\finalYearTrainingSet$, inclusive).

\subsection{Reduce dimensions\label{sec:reduce_dimensions}}

Next we reduce dimensions using principal components analysis (PCA)\ \cite{Lever2017}. Because the data was centered (see Section\ \ref{sec:center_scale}), PCA is equivalent to doing a truncated singular value decomposition. More insights from PCA applied to this exports data are given next in Sec.\ \ref{sec:more_on_PCA}.

\section{Further analysis of the principal components\label{sec:more_on_PCA}}

\subsection{Correlation between the loading on the first principal component and the Product Complexity Index\label{sec:pc0_pci}}

\begin{figure}[htb]
\begin{center}
\includegraphics{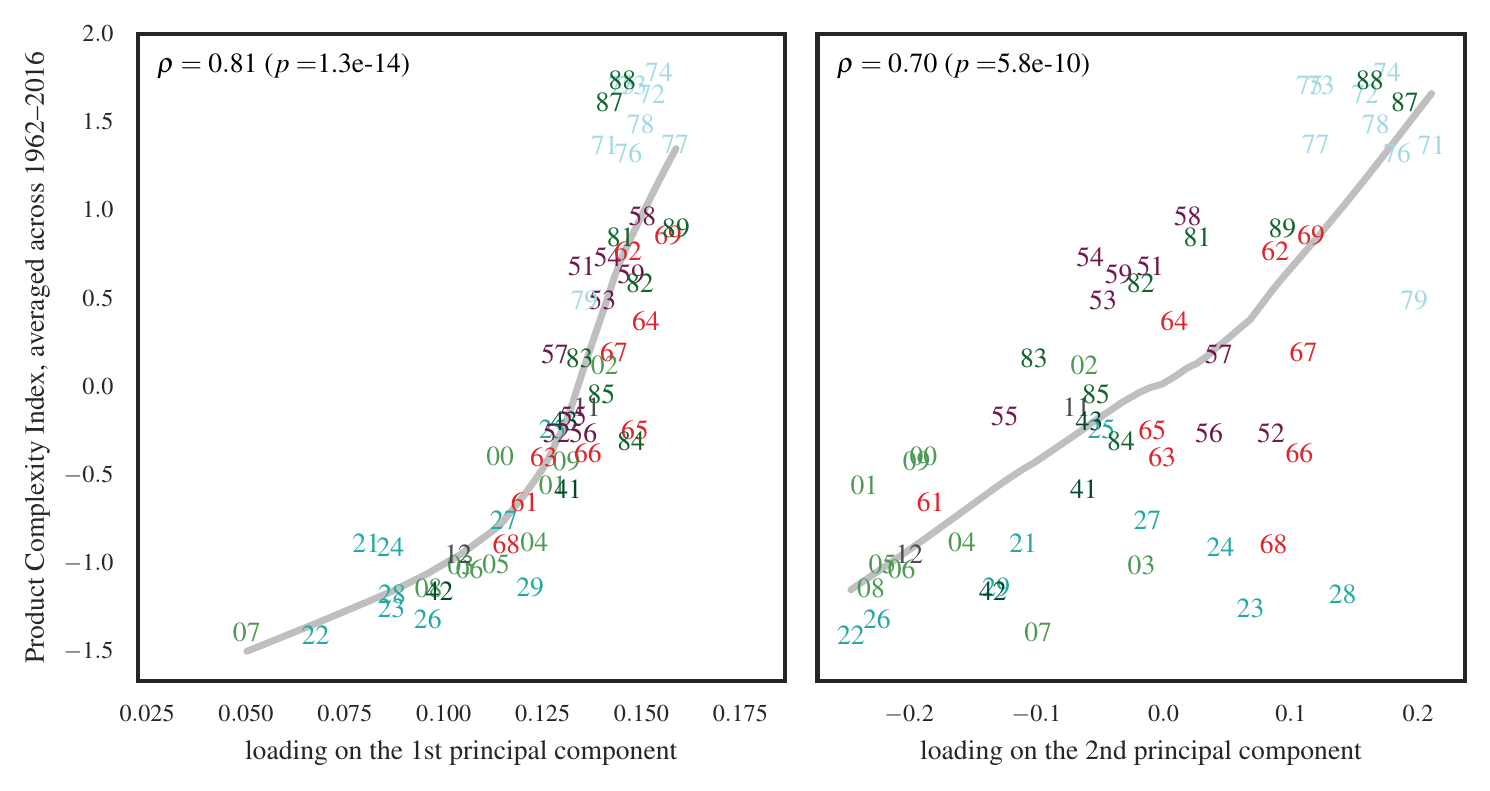}
\caption{
The loadings on the first principal component, and to a lesser degree the loadings on the second principal component, are highly correlated with the Product Complexity Index\ \cite{Hidalgo2009}. In these scatterplots, products are labeled by their 2-digit SITC product codes (available for download \href{https://intl-atlas-downloads.s3.amazonaws.com/index.html}{here}), with colors denoting the first digit. To guide the eye, a locally weighted scatterplot smoothing (LOWESS) is shown in gray; this LOWESS was made using the package \texttt{seaborn} (DOI: \href{10.5281/zenodo.883859}{https://doi.org/10.5281/zenodo.883859}).
}
\label{fig:pci_vs_loading_on_pc0}
\end{center}
\end{figure}

Recall from Fig.\ \ref{fig:pca_loadings} that the first principal component loads positively on all products. But the loadings are not equal: the first principal component loads more on complex products like power generating machinery (product code \texttt{71}) that are produced by few countries, compared to simpler products like vegetables and fruit (product code \texttt{06}) that are produced by many countries. In fact, as shown in Fig.\ \ref{fig:pci_vs_loading_on_pc0}, these loadings are highly correlated with the Product Complexity Index\ \cite{Hidalgo2009}, a notion of complexity (or knowledge intensity) of products based on the complexity (or knowledge intensity) of the countries that produce them. The second principal component is also correlated with the Product Complexity Index, but less so (Pearson correlation $\rho = 0.70$ versus $\rho = 0.81$). 

\subsection{Interpreting a country's score on the first principal component\label{sec:interpret_score_first_pc}}

\subsubsection{Pair-wise correlations\label{sec:interpret_score_first_pc:correlations}}

Figure\ \ref{fig:correlate_score_first_pc_with_others} shows that a country's score on the first principal component, $\scoreFirstPC$, is highly correlated with its total export value per capita [Pearson correlation $\rho = 0.82$, Fig.\ \ref{fig:correlate_score_first_pc_with_others}(D)], which is not surprising given that the loadings of products on the first principal component are all positive. 
However, $\scoreFirstPC$ is more correlated with the Economic Complexity Index\ \cite{Hidalgo2009} [$\rho = 0.82$, Fig.\ \ref{fig:correlate_score_first_pc_with_others}(B)] than is total export value per capita [$\rho = 0.63$, Fig.\ \ref{fig:correlate_score_first_pc_with_others}(E)]. 
The score on the first principal component, $\scoreFirstPC$, is also somewhat correlated with a certain notion of \emph{diversification} of the export basket [$\rho = 0.67$, Fig.\ \ref{fig:correlate_score_first_pc_with_others}(G)]. These observations (together with more reasons given below in Sec.\ \ref{sec:interpret_score_first_pc:regressions}) are why we refer to $\scoreFirstPC$ as ``complexity-weighted diversity''. 
Here, we consider the notion of diversification of exports used in\ \cite[Equation 3]{Hidalgo2009}, namely the number of products $p$ such that the revealed comparative advantage $\RCAsymbol{c}{p}{t}$ exceeds one:
\begin{align}
\diversity{c}{t} := \left \{ p : \RCAsymbol{c}{p}{t} > 1 \right \}.
\label{eq:diversity_RCA}
\end{align}
where 
\begin{align*}
\RCAsymbol{c}{p}{t} \equiv \RCAfraction{c}{p}{t}.
\end{align*}
Figure\ \ref{fig:correlate_score_first_pc_with_others}(G) indicates that export baskets with the highest score $\scoreFirstPC$ on the first principal component
tend to have 
$\RCAsymbol{c}{p}{t}$ larger than one for approximately half of the $59$ $2$-digit products, while the export baskets with the lowest $\scoreFirstPC$ tend to have $\RCAsymbol{c}{p}{t}$ larger than one for fewer than $10$ out of the $59$ $2$-digit products.
Thus, the direction in the space of products in which export baskets over the past $50$ years are most spread out is, loosely speaking, one that distinguishes undiversified, small export baskets from diversified, large ones.

\subsubsection{Intuition behind the correlations\label{sec:interpret_score_first_pc:intuition}}

\begin{figure}[htb]
\begin{center}
\includegraphics[width=\textwidth]{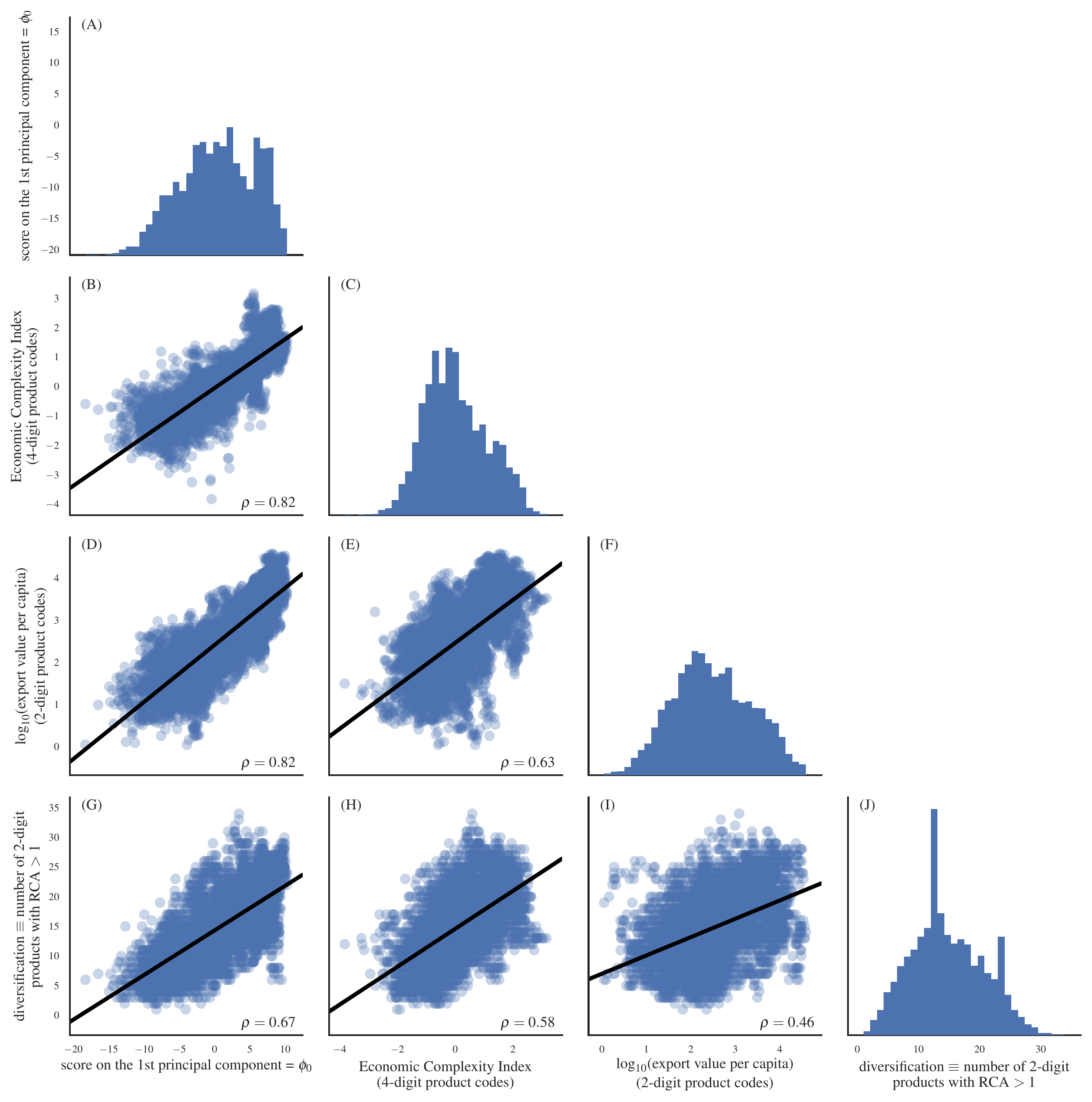}
\caption{
\textbf{The score $\scoreFirstPC$ on first principal component is highly correlated with total per-capita exports [$\rho = 0.82$, panel (D)]; however, compared to total per-capita exports, the score $\scoreFirstPC$ on the first principal component is more correlated with the Economic Complexity Index (ECI) [(B) $\rho = 0.82$ versus (E) $\rho = 0.63$], and it is more correlated with diversification [(G) $\rho = 0.67$ versus (I) $\rho = 0.46$].} 
These differences are one reason why we refer to $\scoreFirstPC$ as ``complexity-weighted diversification''.
This figure shows Pearson correlations ($\rho$) between all pairs of the four variables (1) score $\scoreFirstPC$ on the first principal component, (2) ECI\ \cite{Hidalgo2009}, (3) logarithm (base-10) of total export value per capita, and (4) diversification in terms of revealed comparative advantage (\eqref{eq:diversity_RCA}). 
In the scatterplots, the disks show each (country, year) sample, while the black line shows a least-squares regression. The diagonal shows histograms with 30 bins each. 
The Economic Complexity Index is taken from the same source as the data copied from the \href{http://atlas.cid.harvard.edu}{Atlas at Harvard's Center for International Development} (see Sec.\ \ref{sec:data}) and is computed from product codes at the 4-digit level for all products. All the other data in this figure is from the dataset analyzed in this paper, with countries and products filtered and aggregated at the 2-digit level as described in Sec.\ \ref{sec:filtering}.
}
\label{fig:correlate_score_first_pc_with_others}
\end{center}
\end{figure}

First, we give intuition the correlation between exports per capita and $\scoreFirstPC$ in Fig.\ \ref{fig:correlate_score_first_pc_with_others}(D). 
As shown in Fig.\ \ref{fig:pca_loadings} and\ \ref{fig:pci_vs_loading_on_pc0}, the loadings of the first principal component are positive and range from $0.05$ to $0.15$. This homogeneity of the loadings means that the scores $\scoreFirstPC$ capture an \emph{average scaled absolute advantage}. Given the formula of scaled absolute advantage [\eqref{eq:define_Rpop}, centered and scaled via\ \eqref{eq:center_scale}], we therefore expect $\scoreFirstPC$ to be highly correlated with the logarithm of exports per capita 
[$\rho = 0.82$, Fig.\ \ref{fig:correlate_score_first_pc_with_others}(D)].

For similar reasons, we also expect $\scoreFirstPC$ to be correlated with the diversification of the export basket [Fig.\ \ref{fig:correlate_score_first_pc_with_others}(G)]. To see why, let us drop the index $t$ for clarity of exposition. From the definition of principal component analysis as a singular value decomposition, we get that the matrix $\mathrm{R}$ of absolute advantages can be factored as $\mathrm{R}=\mathrm{H}\mathrm{V}^T$. Here, $H_{ck}$ is the score in the $k$-th principal component for country $c$ (i.e., the $c$-th element in the vector $\scorePC{k}$ in our notation), and $V_{pk}$ is how much product $p$ weights (or loads) on component $k$. The matrix $\mathrm{V}$ is orthogonal, and thus $\mathrm{V}^T\mathrm{V}=\mathrm{I}$ is the identity matrix. Given this decomposition, the 2-norm length of the export vector of country $c$ is 
\begin{align*}
	\left\|R_c\right\|^2 &= [\mathrm{R}\mathrm{R}^T]_{cc}, \\
	&= [\mathrm{H}\mathrm{H}^T]_{cc}, \\
	&= \left\|H_c\right\|^2, \\
	&= \sum_k \scorePC{k}(c)^2.
\end{align*}
In other words, the norm of the export basket vector of country $c$ is equal to the norm of $c$'s vector in the space of principal components. 
Now define a country's diversification in terms of absolute advantage (rather than in terms of RCA as in\ \eqref{eq:diversity_RCA} and \cite{Hidalgo2009}) by discretizing the elements of the vector $R_c$: 
\begin{align*}
	M_{cpt}=
	\begin{cases}
		1,\quad\text{if $R_{cpt}>0$}\\
		0,\quad\text{if $R_{cpt}\leq 0$}
	\end{cases}.
\end{align*}
Having $R_{cpt}>0$ would mean that the country $c$ has an absolute advantage larger than the mean absolute advantage that countries have in that product $p$ in that year $t$. (We could discretize the matrix in other ways, but the result is qualitatively the same.) Then diversity (in terms of absolute advantage) is $d_{c} = \sum_p M_{cp}$, but it is also $d_c = [\mathrm{M}\mathrm{M}^T]_{cc}$. All together, we conclude that when there is a first principal component that explains most of the variation, then
\begin{align}
	d_c \approx \left\|R_c\right\|^2\approx \scoreFirstPC(c)^2 + \scoreSecondPC(c)^2.
	\label{eq:diversity_approx}
\end{align}
Thus, we would expect diversity to be correlated with the square of the first principal component score.

\clearpage

\subsubsection{Regressions of $\scoreFirstPC$\label{sec:interpret_score_first_pc:regressions}}

To investigate whether the score on the first principal component captures information beyond these three quantities Economic Complexity Index, log-exports, and diversification, we use the following datasets:
\begin{description}
	\item[Worldwide Governance Indicators (WGI)] from \url{http://info.worldbank.org/governance/wgi/index.aspx#home}. According to the source, this dataset comprises ``aggregate and individual governance indicators for over 200 countries and territories over the period 1996--2016, for six dimensions of governance: Voice and Accountability, Political Stability and Absence of Violence, Government Effectiveness, Regulatory Quality, Rule of Law, and Control of Corruption.''
	\item[Barro-Lee Educational Attainment Data] from \url{http://barrolee.com/data/Lee_Lee_v1.0/LeeLee_v1.dta} or \url{http://www.barrolee.com/data/BL_v2.2/BL2013_MF1599_v2.2.csv}, which reports ``educational attainment data for 146 countries in 5-year intervals from 1950 to 2010''. It also provides information about the distribution of educational attainment of the adult population over age 15 and over age 25 by sex at seven levels of schooling: no formal education, incomplete primary, complete primary, lower secondary, upper secondary, incomplete tertiary, and complete tertiary. Average years of schooling at all levels---primary, secondary, and tertiary---are also measured for each country and for regions in the world. 
	\item[International Data on Cognitive Skills] from \url{http://hanushek.stanford.edu/sites/default/files/publications/hanushek\%2Bwoessmann.cognitive.xls} which was studied in \cite{Hanushek2012}.
\end{description}
The question is: how much do the quantities and indicators in these datasets explain $\scoreFirstPC$?

\begin{figure}[htb]
\begin{center}
\includegraphics[width=0.7\columnwidth]{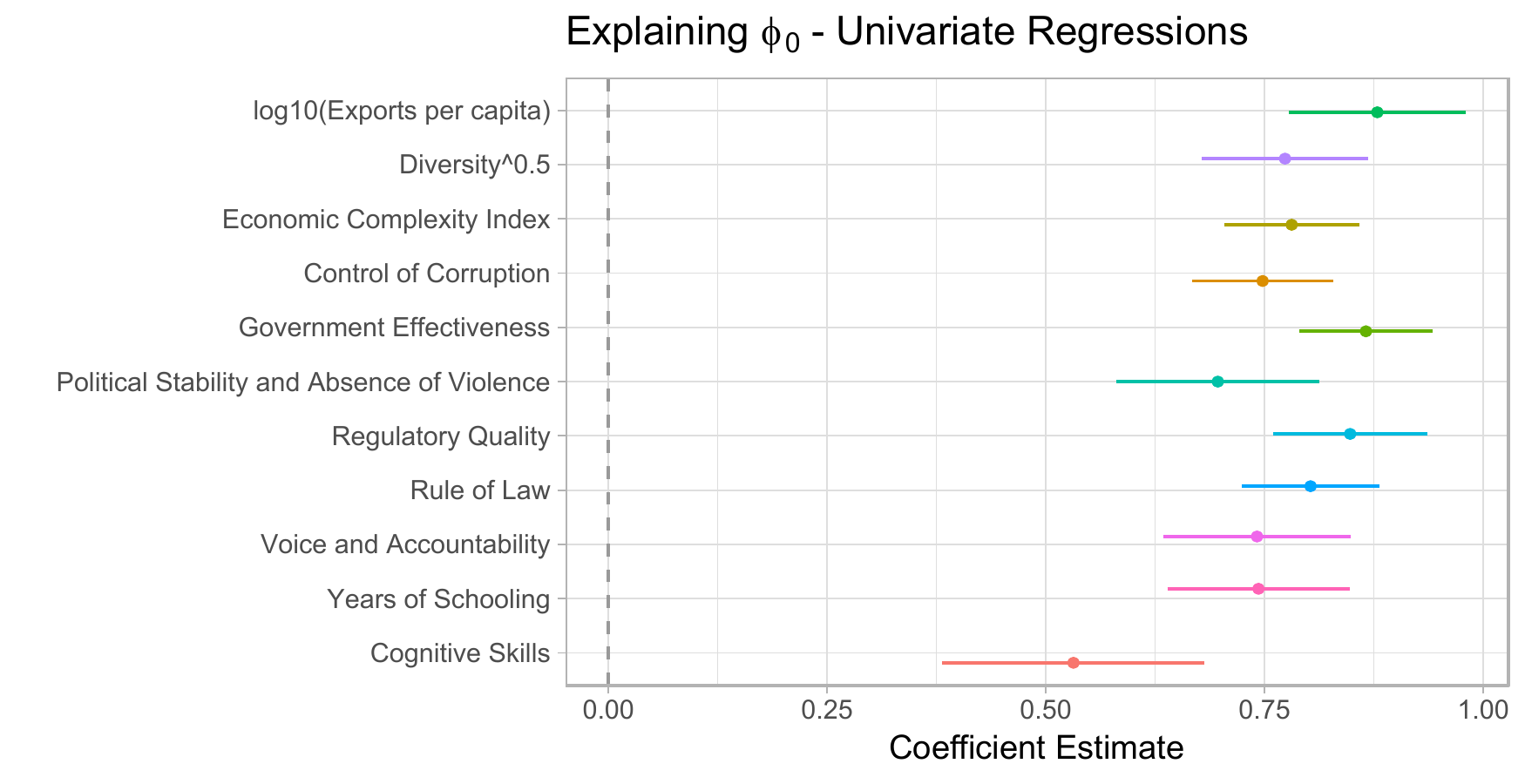}
\caption{
\textbf{Coefficients of the predictors from univariate regressions.} 
 All regressions included year-specific fixed-effects; errors are clustered by country; and error bars reflect $95\%$ confidence intervals. The estimates are for standardized coefficients (i.e., the variables are standardized to have zero mean and unit variance).}
\label{fig:regcoefs_univariate}
\end{center}
\end{figure}

\begin{figure}[htb]
\begin{center}
\includegraphics[width=0.7\columnwidth]{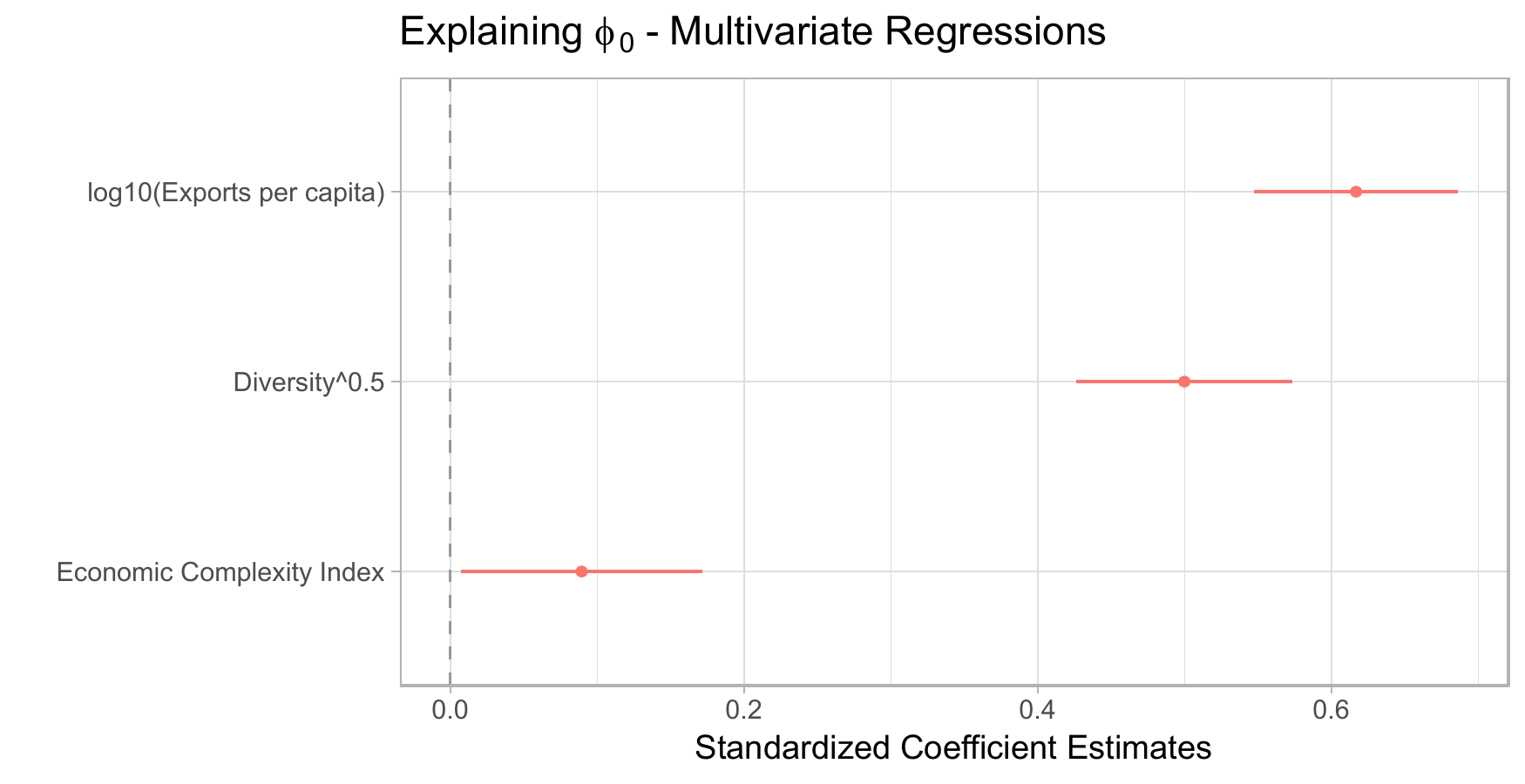}
\caption{
\textbf{Coefficients of the predictors from a multivariate regression only including exports per capita, diversity and economic complexity index.} 
 This regression included year-specific fixed-effects; errors are clustered by country; and error bars reflect $95\%$ confidence intervals. The estimates are for standardized coefficients.}
\label{fig:regcoefs_smallmultivariate}
\end{center}
\end{figure}

\begin{figure}[htb]
\begin{center}
\includegraphics[width=0.7\columnwidth]{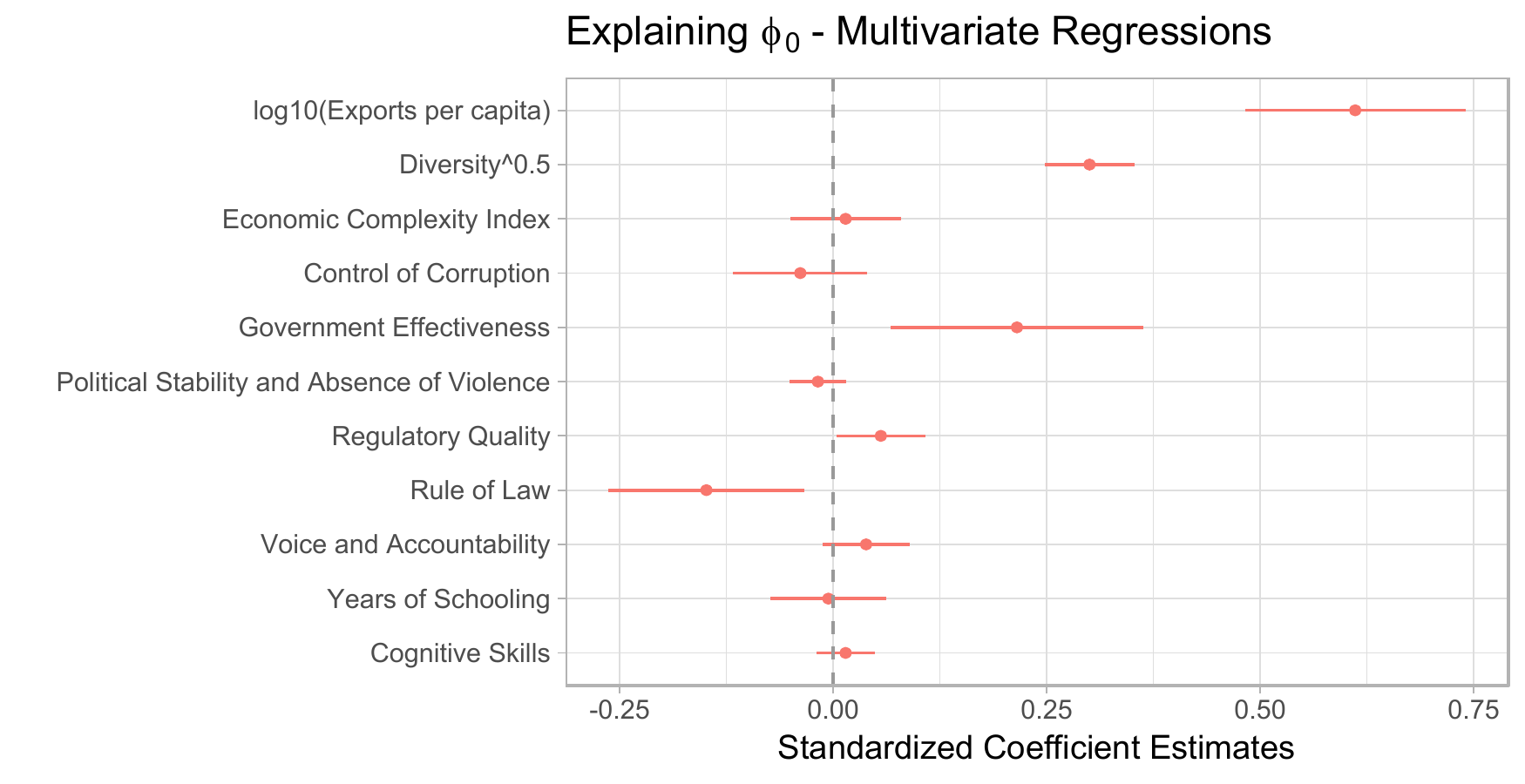}
\caption{
\textbf{Coefficients of the predictors from a multivariate regression including.} 
 This regression included year-specific fixed-effects; errors are clustered by country; and error bars reflect $95\%$ confidence intervals. The estimates are for standardized coefficients.}
\label{fig:regcoefs_multivariate}
\end{center}
\end{figure}

\begin{figure}[htb]
\begin{center}
\includegraphics[width=\columnwidth]{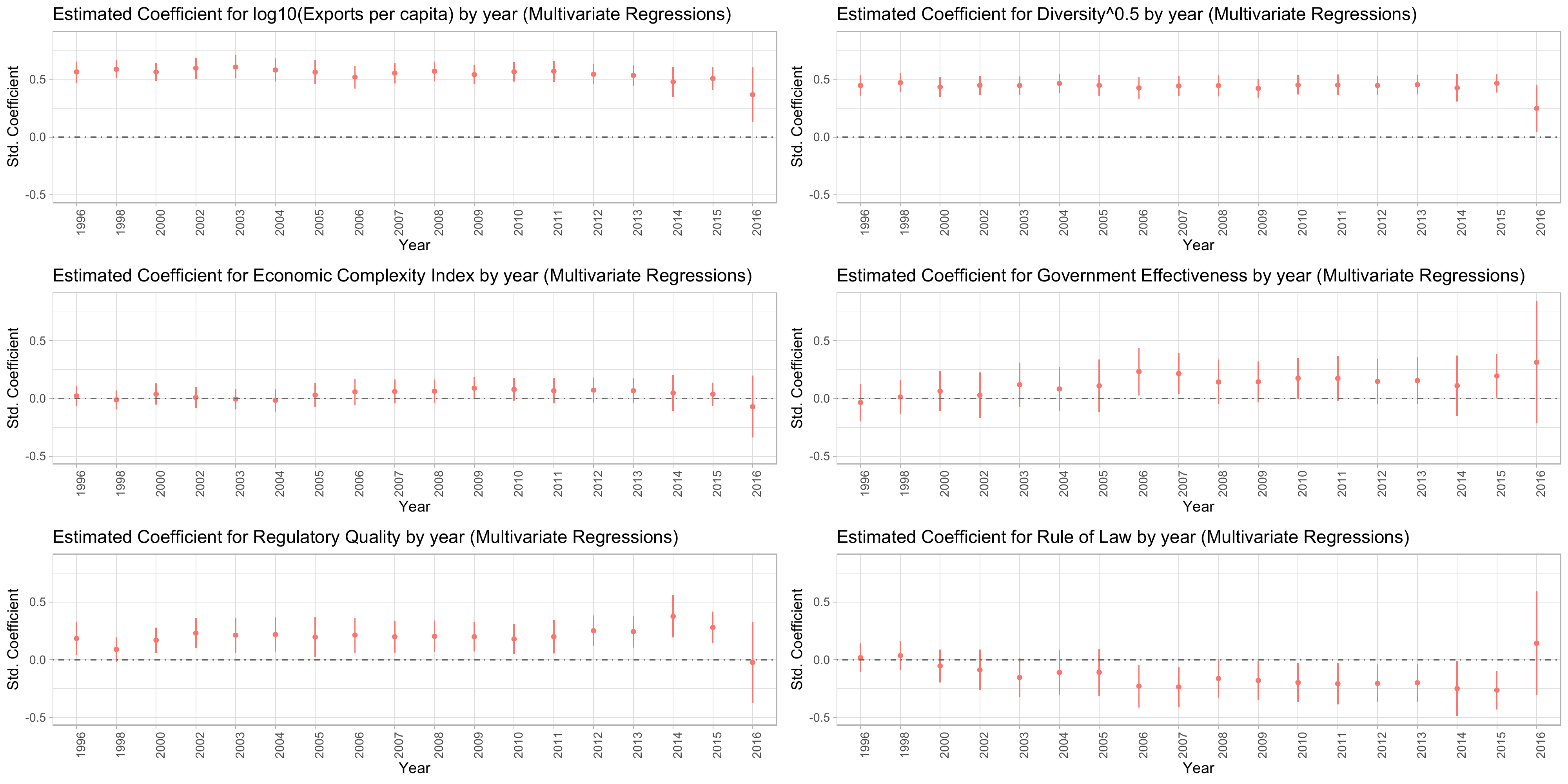}
\caption{
\textbf{Coefficients of the predictors from multivariate regressions carried separately by year.} 
 The estimates are for standardized coefficients.}
\label{fig:regcoefs_acrossyears}
\end{center}
\end{figure}

Figures\ \ref{fig:regcoefs_univariate}, \ref{fig:regcoefs_smallmultivariate}, \ref{fig:regcoefs_multivariate}, and \ref{fig:regcoefs_acrossyears} show the results of the standardized coefficients for different univariate and multivariate regressions. 
In light of the quadratic relationship in\ \eqref{eq:diversity_approx} between diversity (in terms of absolute advantage, $d_c = \sum_p M_{cp}$) and the score on the first principal component, we use the square root of diversity as a predictor of the score on the first principal component. 
While all regressors predict $\scoreFirstPC$ to some extent when we carry out univariate regressions, when all are put together only exports per capita, diversity, government effectiveness, and rule of law survive. In the multivariate regressions done per year, the coefficients for exports per capita and diversity are consistently significant and positive, and have similar magnitudes. In light of these regressions and of the relationship between the loadings on the first principal component with product complexity (Fig.\ \ref{fig:pci_vs_loading_on_pc0}), in the main text we refer to the score $\scoreFirstPC$ on the first principal component as ``complexity--weighted diversity''.

\subsection{More ways of interpreting the first three principal components}
\begin{figure}[htb]
\begin{center}
\includegraphics{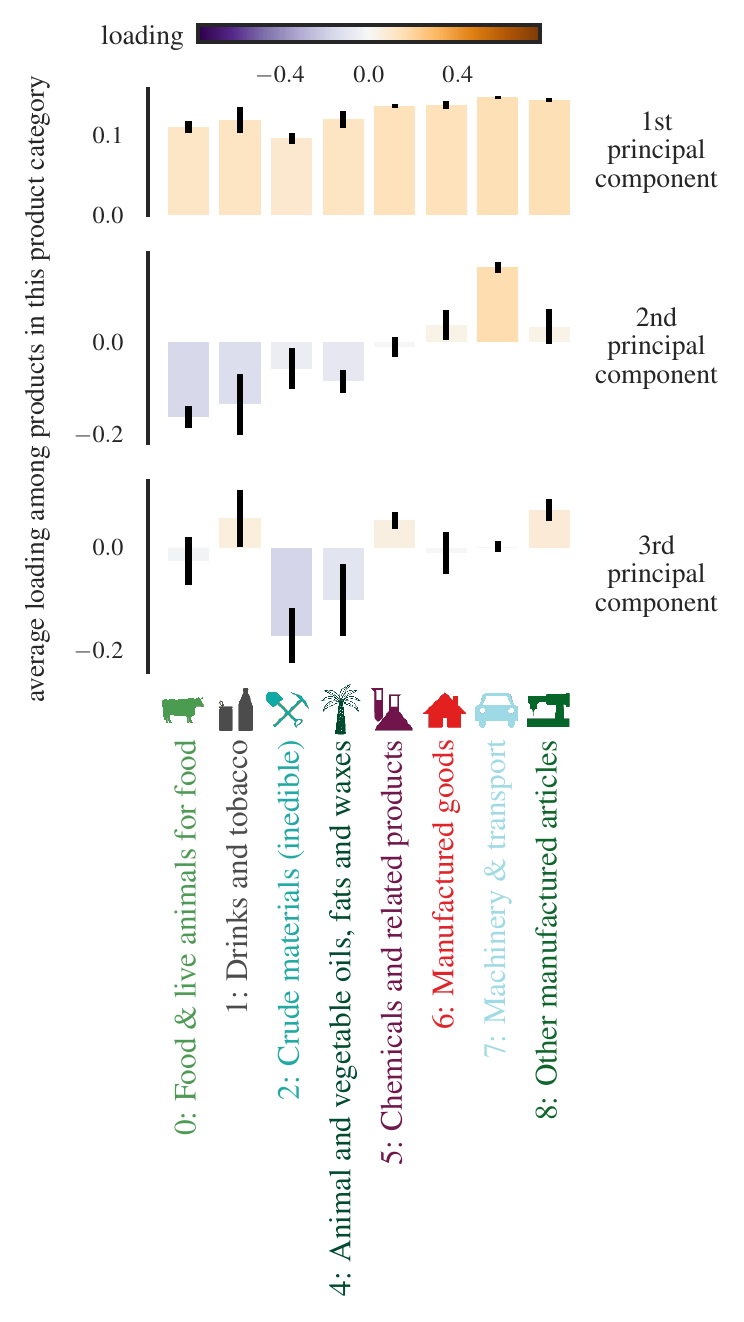}
\caption{PCA loadings grouped and averaged at the 1-digit level.
Error bars show the standard deviation of the loadings divided by the square root of the number of 2-digit products.
}
\label{fig:PCA_loadings_grouped_1digit}
\end{center}
\end{figure}

\paragraph{One-digit product codes}
To help interpret the principal components, in Fig.\ \ref{fig:PCA_loadings_grouped_1digit} we group the $\numProducts$ products at the $1$-digit level and plot the mean and standard deviation. Recall from Sec.\ \ref{sec:filtering} that we removed product category \productCode{3} (i.e., all products with SITC product code that begins with \productCode{3}) to avoid focusing on endowments of fossil fuels.

\begin{figure}[htb]
\begin{center}
\includegraphics[width=\columnwidth]{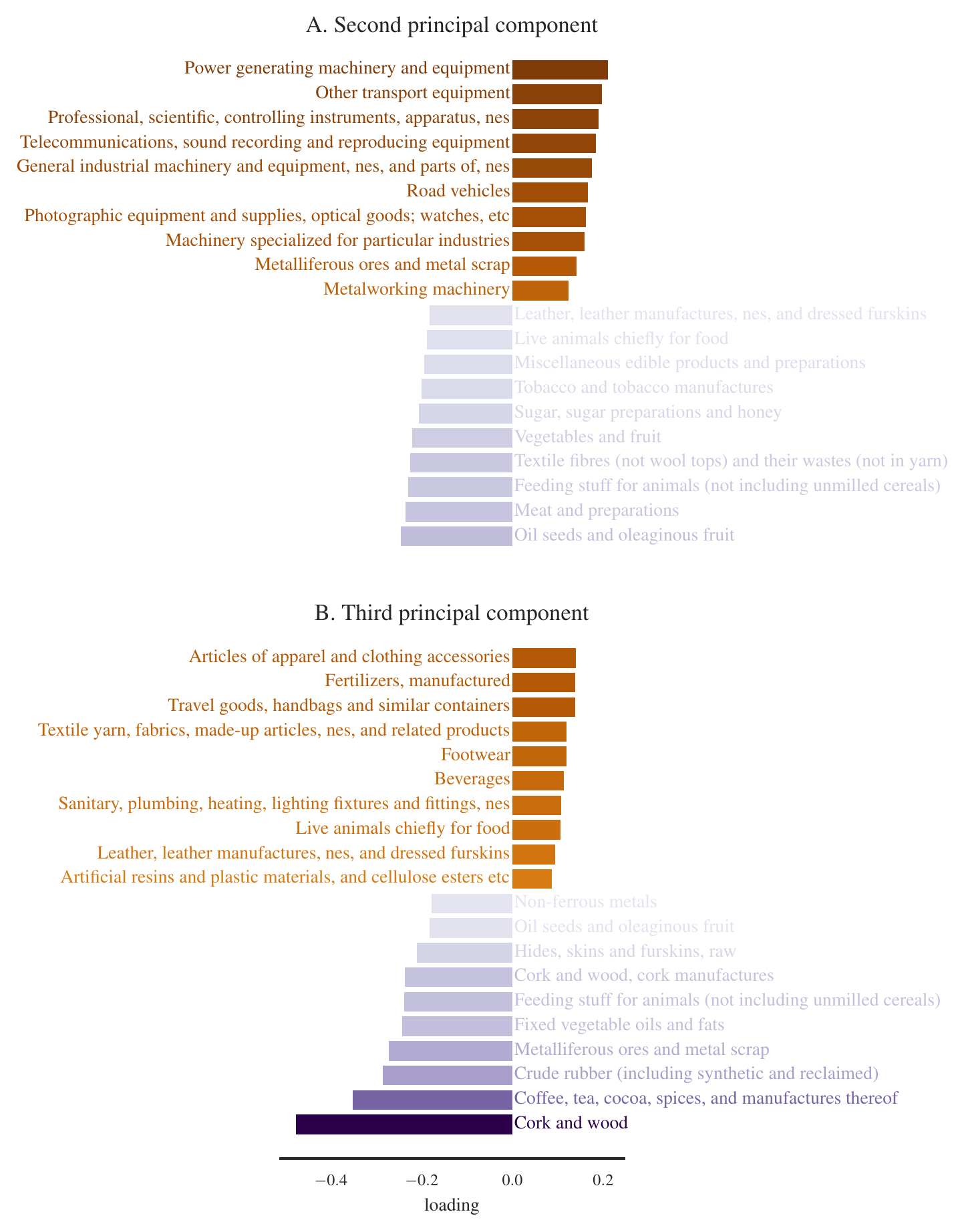}
\caption{Most and least loaded products in the second and third principal components for the data on absolute advantage\ \eqref{eq:define_Rpop}.}
\label{fig:most_least_loaded}
\end{center}
\end{figure}

\paragraph{Most and least loaded 2-digit products in the second and third principal components} 
Figure\ \ref{fig:most_least_loaded} shows 
the top 10 most and least loaded products in the second and third principal components. 
These ``top 10 lists'' aid the interpretation of the first three principal components.

\subsection{Substituting per-capita exports or diversification for the score $\scoreFirstPC$ on the first principal component suggests that diversification, not simply a rise in total exports per capita, precedes economic growth\label{sec:substitute_for_score_first_pc}}

Rich countries are usually big exporters, have diversified economies, and produce complex products. Hence, these quantities correlate positively with each other, which makes it particularly difficult to interpret the meaning of scores resulting from PCA. We find in the main text that high levels in $\scoreFirstPC$ precede growth in income. But the scores of $\scoreFirstPC$ are correlated with both exports per-capita and product diversification, so this relationship could mean either that high levels of export per-capita precede growth, or that high levels of diversification precede growth, or both. To better understand what $\scoreFirstPC$ represents in our analysis and what it reveals about economic development, here we substitute another variable for it in the GAM: either per-capita export value, or another definition of ``product diversification''.

\subsubsection{Exports per capita are less strongly associated with growth in incomes compared to $\scoreFirstPC$}

\begin{figure}[htb]
\begin{center}
\includegraphics[width=\columnwidth]{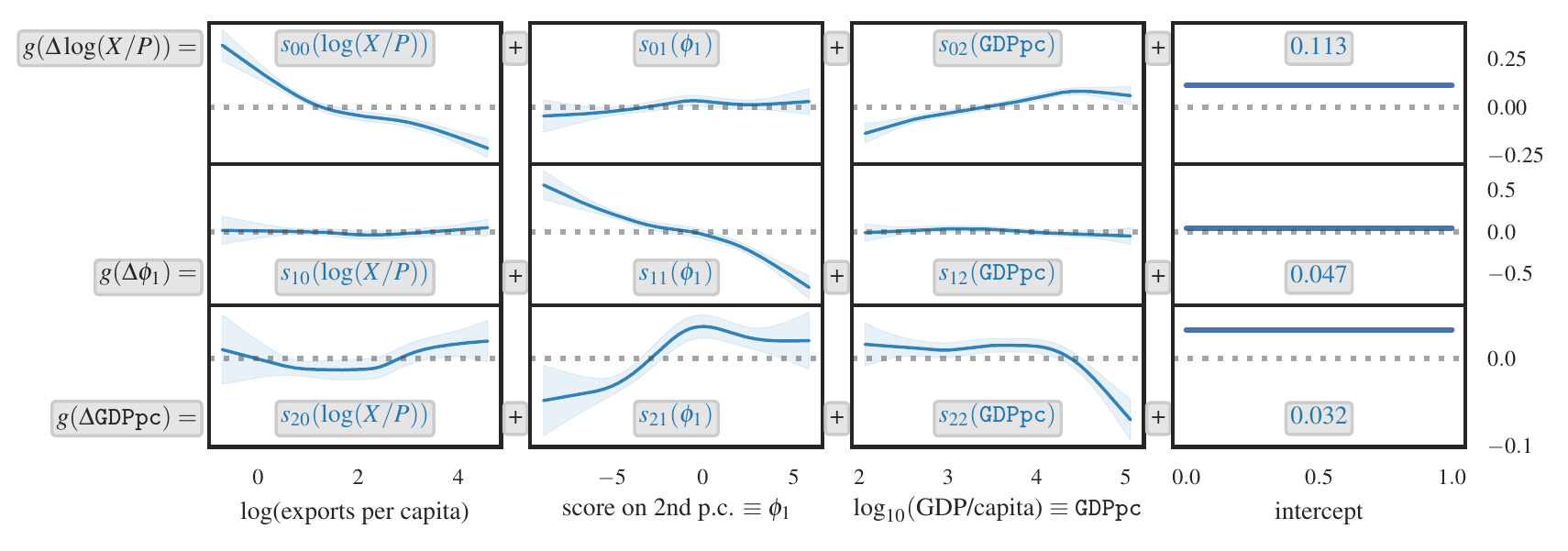}
\caption{Substituting total export value per capita $X / \populationSymbol$ for $\scoreFirstPC$ results in a flat relationship between $X / \populationSymbol$ and growth in income (bottom-left plot). Compare this partial dependence with Fig.\ \ref{fig:partial_dependence} and Fig.\ \ref{fig:partial_dependence_diversification}.}
\label{fig:partial_dependence_exports_per_capita}
\end{center}
\end{figure}

Figure\ \ref{fig:partial_dependence_exports_per_capita} shows a partial dependence plot (like Fig.\ \ref{fig:partial_dependence}) from a model fitted to the same dataset except that $\scoreFirstPC$ is replaced by total export value per capita. Note in particular the bottom-left plot: The 95\% confidence interval of the relationship between economic growth and export value per capita contains zero or is slightly below zero, suggesting a weak relationship between rises exports (no matter the product) and economic growth. Contrast this flat relationship with the positive trend in the bottom-left plot of Fig.\ \ref{fig:partial_dependence}.

\subsubsection{Replacing $\scoreFirstPC$ with another notion of diversification results in qualitatively similar results}

\begin{figure}[htb]
\begin{center}
\includegraphics[width=\columnwidth]{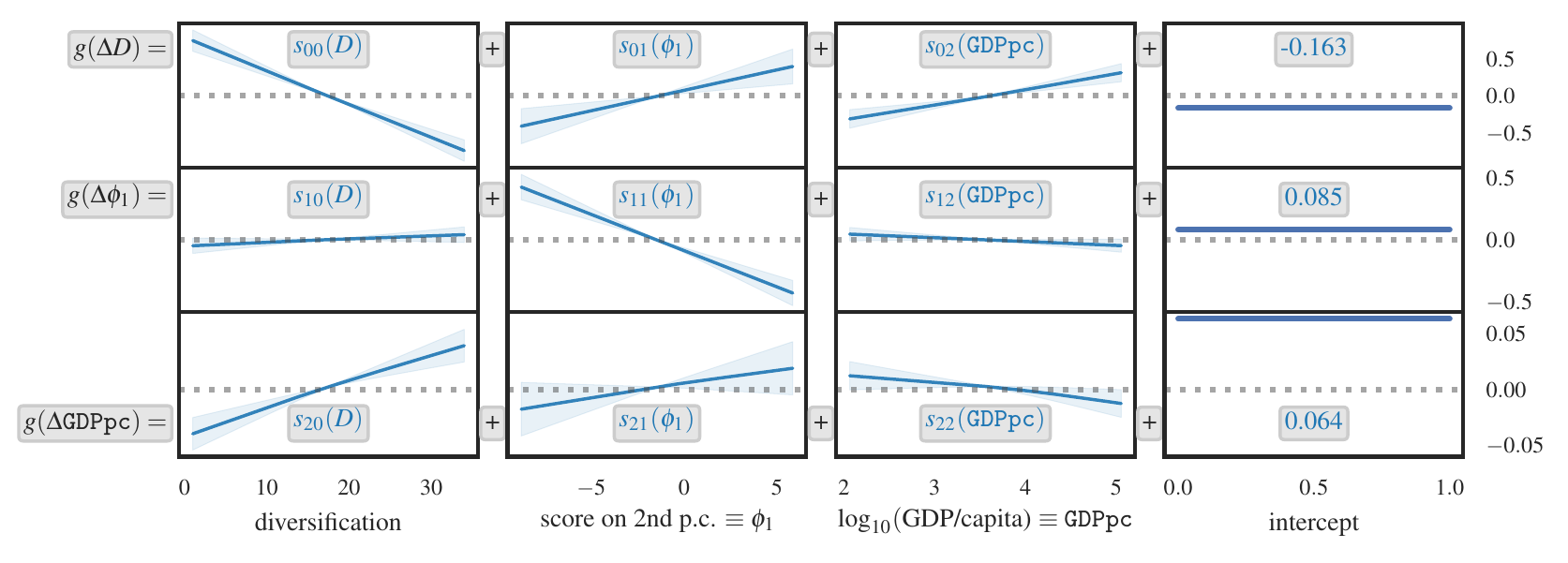}
\caption{Replacing $\scoreFirstPC$ with the definition of diversification defined in Eq. [3] in \cite{Hidalgo2009} (the number of products with revealed comparative advantage (RCA) larger than one) results in a qualitatively similar model that is more linear than the one discussed in the main text. Compare this partial dependence with Fig.\ \ref{fig:partial_dependence} and Fig.\ \ref{fig:partial_dependence_exports_per_capita}.}
\label{fig:partial_dependence_diversification}
\end{center}
\end{figure}

Next we tried replacing $\scoreFirstPC$ by diversification as defined in Eq. [3] in \cite{Hidalgo2009}: the number of products with revealed comparative advantage (RCA) larger than one. 
The resulting GAM is approximately linear and behaves qualitatively similarly to the model in the main text; in fact, it appears to be a linear approximation of that GAM. 
In light of this resemblance to the model in the main text, it seems reasonable to call $\scoreFirstPC$ something akin to diversification; here, we call $\scoreFirstPC$ ``complexity-weighted diversification''. The comparison between Figs.\ \ref{fig:partial_dependence},\ \ref{fig:partial_dependence_exports_per_capita}, and\ \ref{fig:partial_dependence_diversification} suggests that exporting a large diversity of complex products precedes economic growth.

\section{Details about the generalized additive model: Training and performance\label{sec:gam_training_performance}}

\subsection{Cubic smoothing splines using the B-spline basis\label{sec:pyGAM}}
Generalized additive models were estimated using the package \href{https://pypi.python.org/pypi/pygam/0.2.17}{\texttt{pyGAM 0.2.17}}\ \cite{pyGAM}, which uses a B-spline basis, computed using De Boor recursion. The basis functions extrapolate linearly past the end-knots. Details on cubic smoothing splines are in\ \cite[Chapters 3 and 4]{Wood2006book} and\ \cite[Sec. 5.4]{ESLbook}. 

\subsubsection{Nested-in-time cross validation\label{sec:cross_validation}}
The generalized additive model (GAM)\ \eqref{eq:gam} has two hyperparameters: the smoothing strength $\lambda$ that penalizes wiggliness, and the number of splines. Following the advice of\ \cite{Wood2006book}, we tried relatively large values for the number of splines (uniformly distributed over $\{15, 16, \dots, 60\}$) and let the smoothing penalty do the regularization. We sampled $\log_{10} \lambda$ uniformly over $[-3.0, 10.0]$.

To choose the best hyperparameters, we split the data into five training sets that are nested in time as follows. 
The task is to predict the change in the time-series between year $t-1$ and $t$ given the value of the time-series at year $t-1$ (i.e., autoregression with lag $1$). 
We put the earliest $39\%$ of samples in the first training set, and then we partition the remaining samples into roughly equal-size sets. 
(Because countries appear and disappear, some care needs to be taken with time-series of different lengths; we use quantiles of the times of all the samples to find where to split the data.) 
The result is that the first training set is data with $t$ between $\beginningYearTrainingSet$ and $\finalYearTrainingSet$, and the corresponding test set is data with $t$ between $1989$ and $1995$. The train--test splits are shown in Table\ \ref{tab:cv_splits}.
\begin{table}[htp]
\caption{Train--test splits used in cross validation, and coefficient of determination ($R^2$) averaged across the three prediction problems (predict annual changes in $\scoreFirstPC$, $\scoreSecondPC$, and $\gdppc$) for the GAM\ \eqref{eq:gam} and a baseline model that predicts the average change observed in the training set. The model's task is to predict year $t$ using data about year $t-1$.}
\begin{center}
%
\begin{tabular}{clcc}
\toprule
{} & {} & \multicolumn{2}{c}{Average $R^2$ across $3$ targets}\\
{} & {} & GAM\ \eqref{eq:gam} & Baseline model\\
Split & {} & {} & {}  \\
\midrule
\multirow{2}{*}{0} 
& Train $1963 \leq t \leq 1988$ &  $0.038$ &    $-0.002$ \\
& Test $1989 \leq t \leq 1995$ &  $0.044$ &    $-0.034$ \\
\multirow{2}{*}{1} 
& Train $1963 \leq t \leq 1995$ &  $0.045$ &    $-0.005$ \\
& Test $1996 \leq t \leq 2001$ &  $0.019$ &    $-0.021$ \\
\multirow{2}{*}{2} 
& Train $1963 \leq t \leq 2001$ &  $0.046$ &    $-0.002$ \\
& Test $2002 \leq t \leq 2006$ & $-0.083$ &    $-0.102$ \\
\multirow{2}{*}{3} 
& Train $1963 \leq t \leq 2006$ &  $0.044$ &    $-0.002$ \\
& Test $2007 \leq t \leq 2011$ &  $0.000$ &    $-0.010$ \\
\multirow{2}{*}{4} 
& Train $1963 \leq t \leq 2011$ &  $0.040$ &    $-0.002$ \\
& Test $2012 \leq t \leq 2016$ & $-0.135$ &    $-0.167$ \\
\midrule
\multirow{2}{*}{Average}
& Train & $0.042$ &    $-0.002$ \\
& Test &  $-0.031$ &    $-0.067$ \\
\bottomrule
\end{tabular}
\end{center}
\label{tab:cv_splits}
\end{table}

The model is always tested on data from the future relative to the test set. With this cross validation scheme, the hyperparameters with best performance on the test sets were smoothing strength $\lambda = 2748.5$ and $37$ splines. These values were used for each of the three equations in equation\ \eqref{eq:gam}.

\subsubsection{The GAM outperforms a baseline model that predicts the average change in the test set\label{sec:performance_R2}}

\begin{figure}[htb]
\begin{center}
\includegraphics[width=.95 \textwidth]{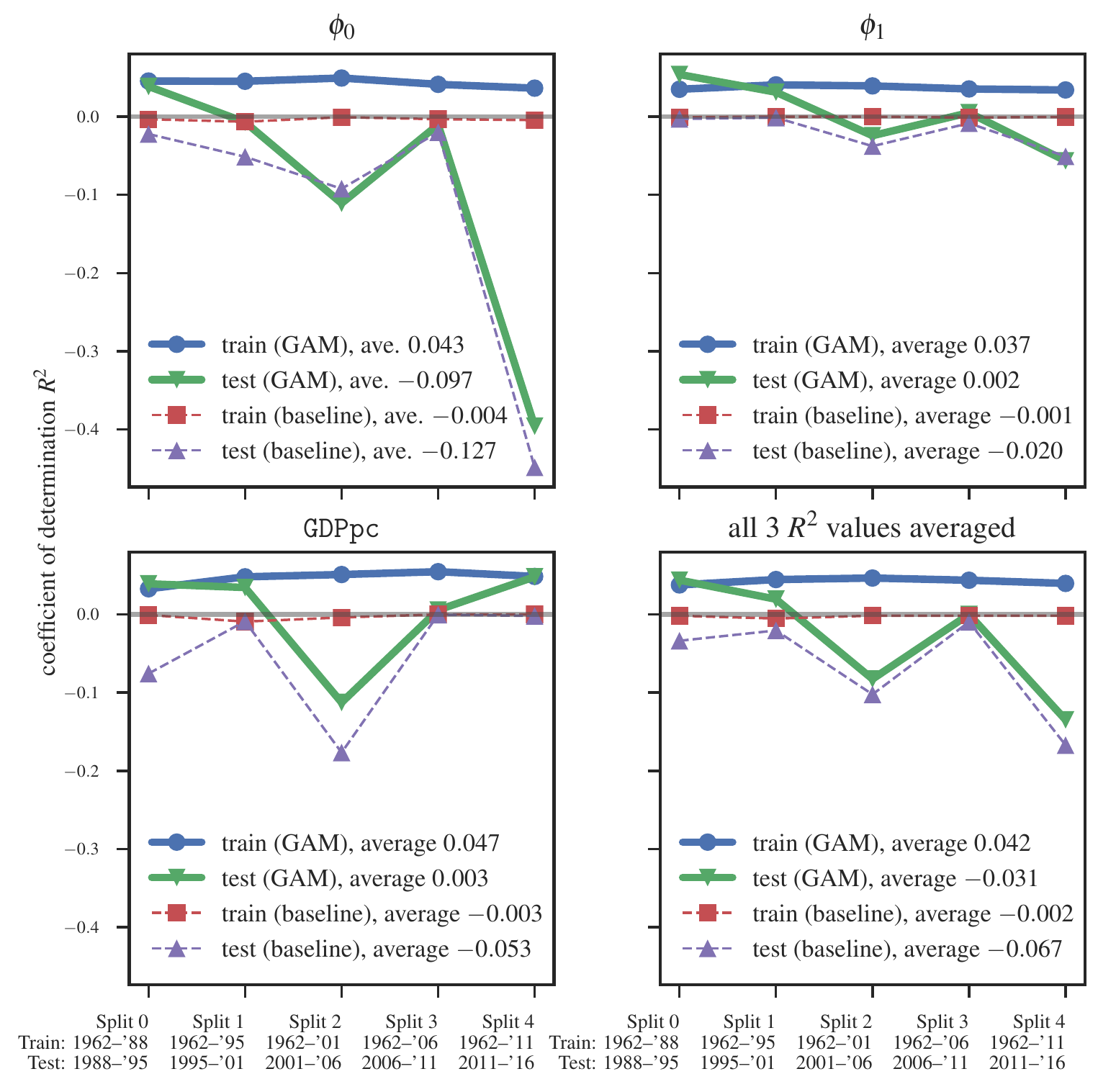}
\caption{\textbf{The GAM\ \eqref{eq:gam} outperforms a baseline model that predicts the average change seen in the test set.} However, the performance of the GAM on the test set has considerable room for improvement: for all three prediction problems of predicting $\scoreFirstPC$, $\scoreSecondPC$, and $\gdppc$, the $R^2$ is approximately $0.04$ on the training set and slightly below zero on the test set.}
\label{fig:R2}
\end{center}
\end{figure}

Table\ \ref{tab:cv_splits} and Fig.\ \ref{fig:R2} show the performance of the GAM\ \eqref{eq:gam} in terms of the coefficient of determination, $R^2$. The GAM does better than a simple baseline model that simply predicts the average change observed in the training set. (This baseline model predicted the test set more accurately than a baseline model that predicted the median of the training set.) However, the performance of the GAM has considerable room for improvement: its $R^2$, averaged across the three prediction problems (of predicting $\scoreFirstPC$, $\scoreSecondPC$, and $\gdppc$) is $0.042$ on the training sets and $-0.031$ on the test sets. For some reason, the score $\scoreFirstPC$ is particularly difficult to predict in the most recent test set ($2011$--$2016$, in ``Split 4''). Also, per-capita incomes were difficult to predict in the test set of Split 2 ($2001$--$2006$).

We tried several alternative modeling strategies other than GAMs, including neural networks, random forests, and kernel ridge regression. None of these competing methods significantly outperformed GAMs in their accuracy on test sets, and they were less readily interpretable than GAMs, so we chose to focus on GAMs. That more flexible modeling strategies could not significantly outperform GAMs, despite our best efforts at searching over a large set of hyperparameters, indicates just how difficult it is to predict the dynamics of national economies.

\subsection{Quantile--quantile plots and raising the response variable sto the $\linkRoot$ power\label{sec:QQplots}}

\begin{figure}[htb]
\begin{center}
\includegraphics[width=\textwidth]{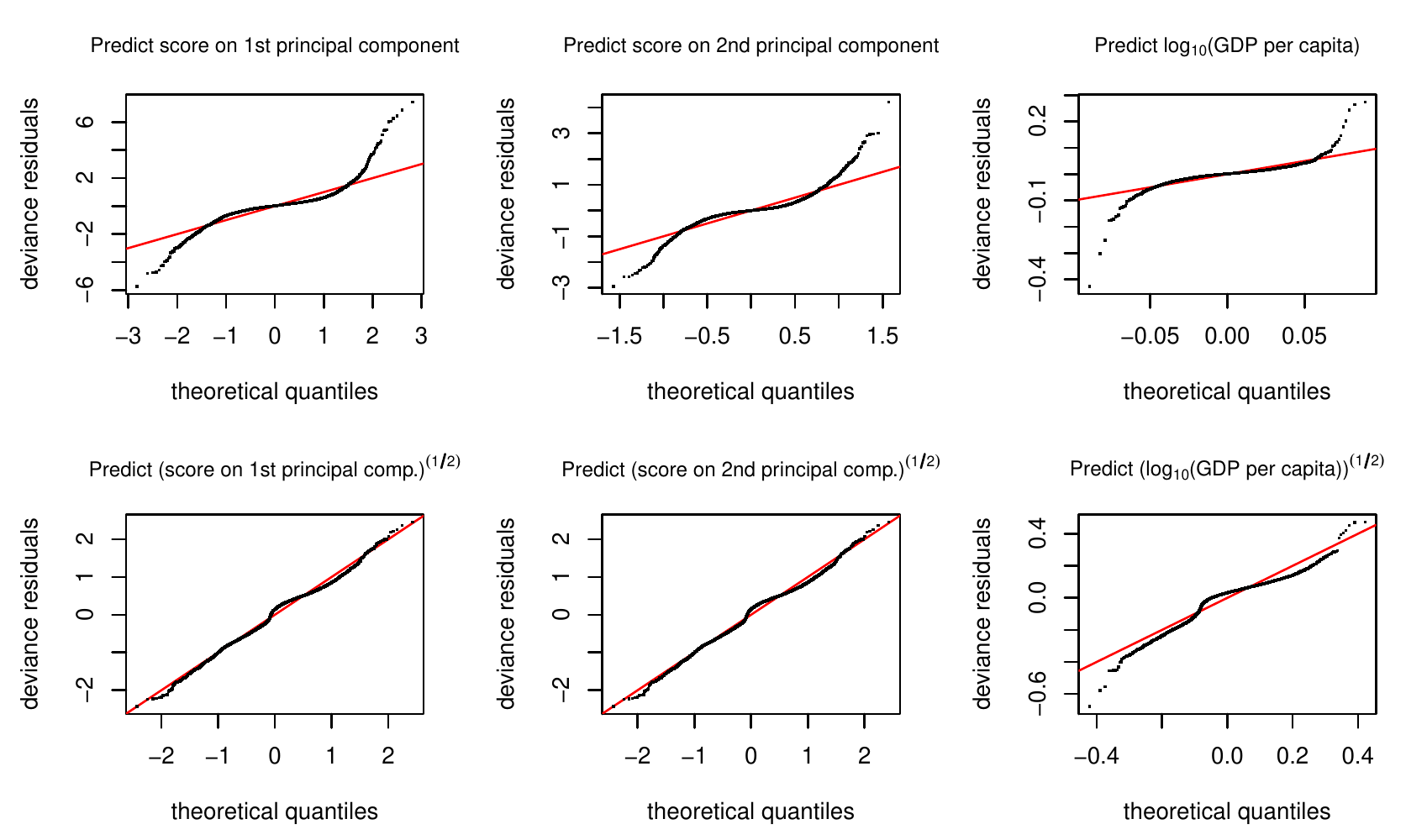}
\caption{Quantile-quantile plots for the task of predicting yearly changes $\scoreFirstPC(t+1) - \scoreFirstPC(t)$, $\scoreSecondPC(t+1) - \scoreSecondPC(t)$, $\gdppc(t+1) - \gdppc(t)$ (top row, left to right) and for predicting those yearly transformed by $\link{x} \equiv \linkExpression{x}$ (bottom row).
In the top row, we see significant improvement in how close the deviance residuals are to the straight red line. 
}
\label{fig:QQ_Rpop}
\end{center}
\end{figure}

Figure\ \ref{fig:QQ_Rpop} shows a quantile-quantile (QQ) plot of the residuals of generalized additive models of the form\ \eqref{eq:gam} with the response transformed by $\link{x} \equiv \linkExpression{x}$ (bottom row) or not (top row). The residuals are conditioned on the fitted model coefficients and scale parameter. The closer the QQ-plot is to a straight line, the better the distributional assumptions are satisfied. The QQ-plots were made using the function \texttt{qq.gam} in the package \texttt{mgcv} by Simon Wood\ \cite{Wood2011fast}. 

Because the QQ-plots are closer to a straight line when we transform the response with $\link{x} \equiv \linkExpression{x}$ (compare bottom row and top row in Fig.\ \ref{fig:QQ_Rpop}), we transform the response variables with the invertible function $\link{x} \equiv \linkExpression{x}$. (An example of such a transformation based on the results of \texttt{qq.gam} is given on page 230 in Section 5.2.1 in\ \cite{Wood2006book}.) When making iterated predictions (as in Figure\ \ref{fig:predicted_growth_rates}), we invert $\link{x}$ in order to feed the response back into the model as a predictor variable.

\subsection{Errors averaged by country\label{sec:errors_by_country}}

\begin{figure}[htb]
\begin{center}
\includegraphics[width=\textwidth]{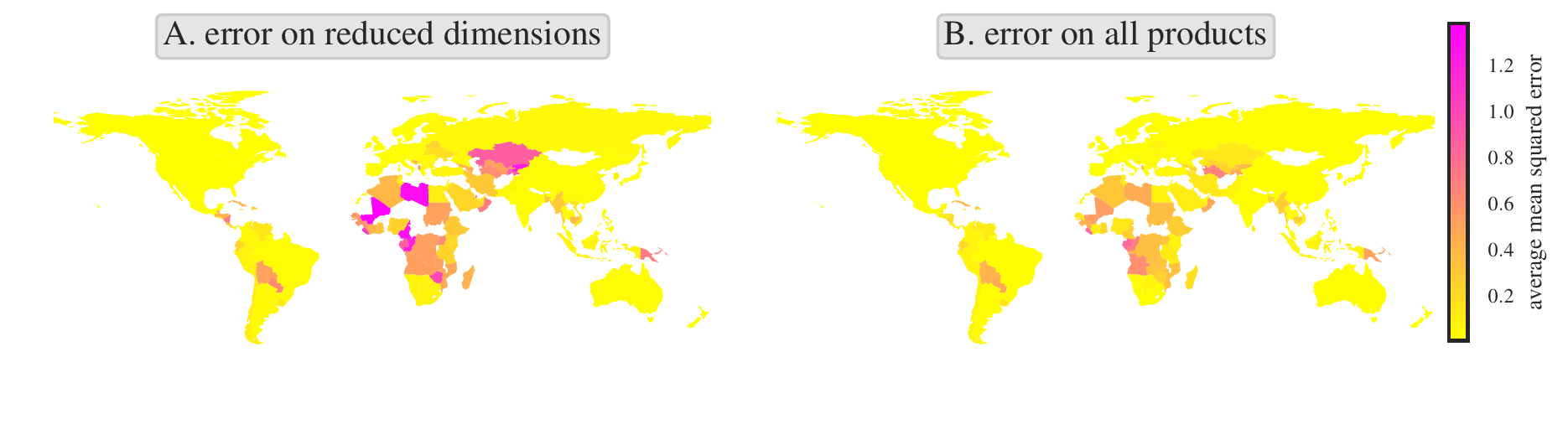}
\caption{
\textbf{
The inferred model tends to make larger errors in predicted changes of export baskets and per-capita incomes of low-income countries, especially in Africa.
}
	Plotted are the squared errors in predicting export baskets and per-capita incomes, averaged across columns [i.e., across $(\scoreFirstPC, \scoreSecondPC, \gdppc)$] and then averaged across time.
The left-hand column shows the error on the reduced dimensions $(\scoreFirstPC, \scoreSecondPC, \gdppc)$, while the right-hand column shows the errors after the principal component scores $(\scoreFirstPC, \scoreSecondPC)$ are inverted back to the original dimensions corresponding to $\numProducts$ products.}
\label{fig:errors_averaged_by_country}
\end{center}
\end{figure}

This model tends to be more accurate for more developed countries, as shown in Figure\ \ref{fig:errors_averaged_by_country}. 
Figure\ \ref{fig:errors_averaged_by_country} shows the squared errors averaged by country, for predicting export baskets that have been dimension-reduced with PCA (left column) and for predicting the export baskets themselves (right-column).

The trajectories of poorer countries in Figures\ \ref{fig:streamplot_scatterplot_pc0_p1_and_pc0_gdppc} and\ \ref{fig:streamplots_vary_gdppc} in the main text appear to be laminar. By contrast, Cristelli et al.\ \cite{Cristelli2015} found that the trajectories of the poorest countries are turbulent when they analyzed yearly changes in ``fitness'' and per-capita incomes. The dynamics in our model\ \eqref{eq:gam} are laminar because a large smoothing strength is chosen in cross validation (Sec.\ \ref{sec:cross_validation}). However, the greater predictability of richer countries (Fig.\ \ref{fig:errors_averaged_by_country}) is consistent with the finding of Cristelli et al.\ \cite{Cristelli2015} that richer countries move in a more laminar, predictable path through the space defined by per-capita income and by a summary measure of export baskets.

\subsection{Alignment of changes in export baskets with the gradient of per-capita incomes}

\subsubsection{Countries tend to ``hill climb'' to higher incomes}

\begin{figure}[htb]
\begin{center}
\includegraphics{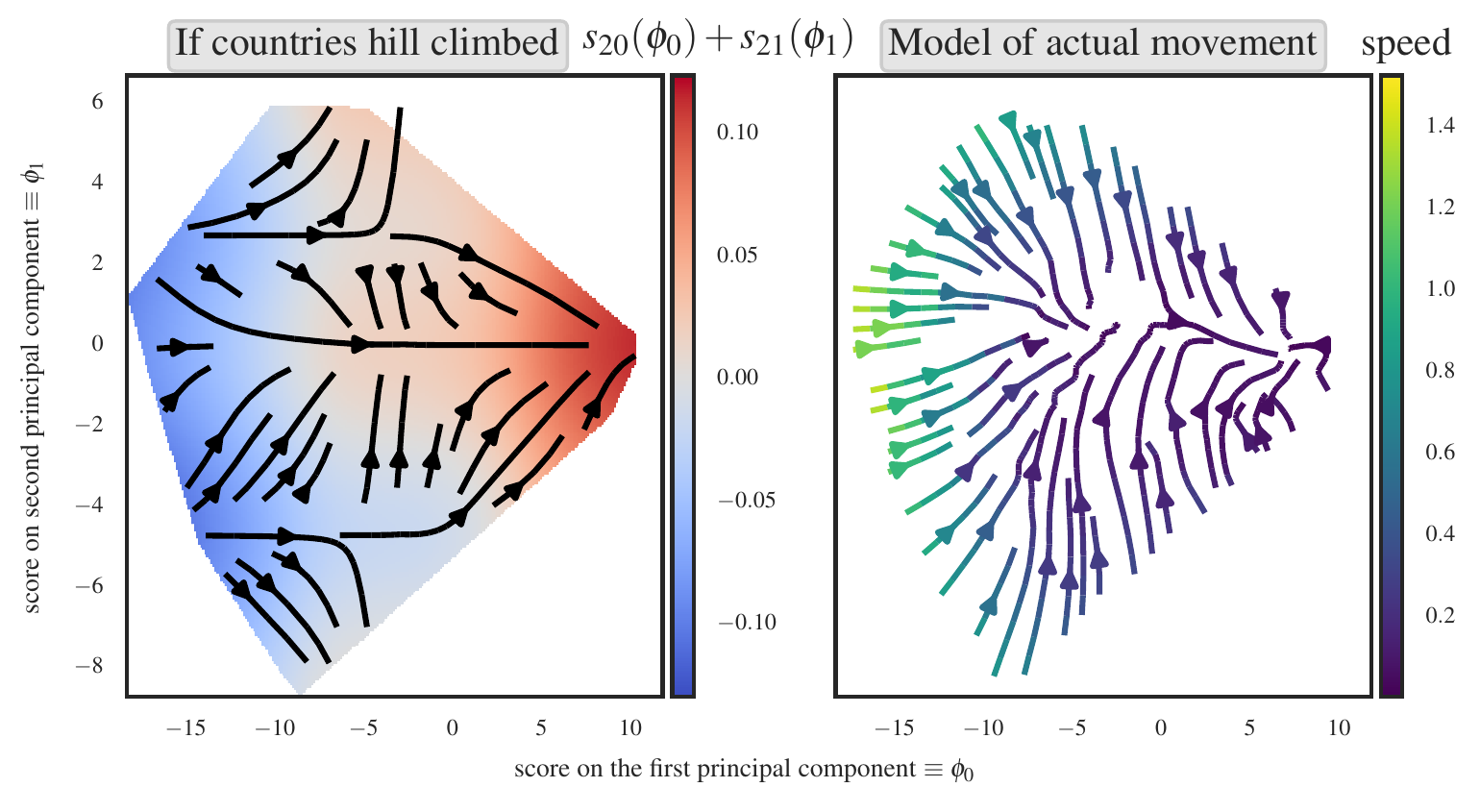}
\caption{
\textbf{
Hill climbing: countries tend to 
change their export baskets to maximize short-run gains in per-capita incomes.}
In the left column, the contribution of (dimension-reduced) export baskets to the changes in per-capita incomes, $\smoothTerm_{20}(\scoreFirstPC) + \smoothTerm_{21}(\scoreSecondPC)$, is plotted using colors in a blue--red spectrum. 
The black streamlines show the gradient of that mapping; they mark the direction in which a country would change its $\scoreFirstPC$ (roughly speaking, its export diversity) and $\scoreSecondPC$ (roughly speaking, its exports of agriculture minus machinery) to maximize next year's per-capita income, according to the fitted model. 
The right column shows a smoothed version of how countries actually move through $(\scoreFirstPC, \scoreSecondPC)$, with colors denoting the speed of movement, $\sqrt{\left ( \Delta \scoreFirstPC \right ) ^2 + \left ( \Delta \scoreSecondPC \right )^2}$. 
For each rectangle in a fine grid of rectangles covering the diagram, we find the per-capita income $\widetilde g$ of the sample with closest $(\scoreFirstPC, \scoreSecondPC)$ to a corner $(\widetilde \scoreFirstPC, \widetilde \scoreSecondPC)$ of that rectangle, and then we plot the predicted movement in $(\scoreFirstPC, \scoreSecondPC)$ evaluated at  $(\widetilde \scoreFirstPC, \widetilde \scoreSecondPC, \widetilde g)$ according to the cubic-spline model\ \eqref{eq:gam}. 
%
}
\label{fig:hill_climb}
\end{center}
\end{figure}

Do economies' export baskets change in ways that lead to rising incomes? 
To explore that question, 
we plot in the left column of Figure\ \ref{fig:hill_climb} the direction in $(\scoreFirstPC, \scoreSecondPC)$ that would most increase per-capita incomes [i.e., the gradient $\left (\smoothTerm_{20}'(\scoreFirstPC), \smoothTerm_{20}'(\scoreSecondPC) \right)$. 
For comparison, in the right-hand column of Figure\ \ref{fig:hill_climb} we plot the typical movement in $(\scoreFirstPC, \scoreSecondPC)$ according to the fitted model\ \eqref{eq:gam}. In these plots on the right-hand column, at each point in a fine grid of points, we find the $\gdppc$ of the closest sample to that grid point. 
This procedure results in more wiggles in the streamlines compared to when $\gdppc$ is fixed at a certain value, as in Figure\ \ref{fig:streamplots_vary_gdppc} in the main text.

By comparing the left- and right-hand plots in Figure\ \ref{fig:hill_climb}, we see how well economies tend to ``hill climb'' toward higher per-capita incomes according to the model\ \eqref{eq:gam}. 
Except for two extreme points where few observations are found (very high and very low $\scoreSecondPC$), countries do tend to move along the gradient of per-capita income.

\begin{figure}[htb]
\begin{center}
\includegraphics{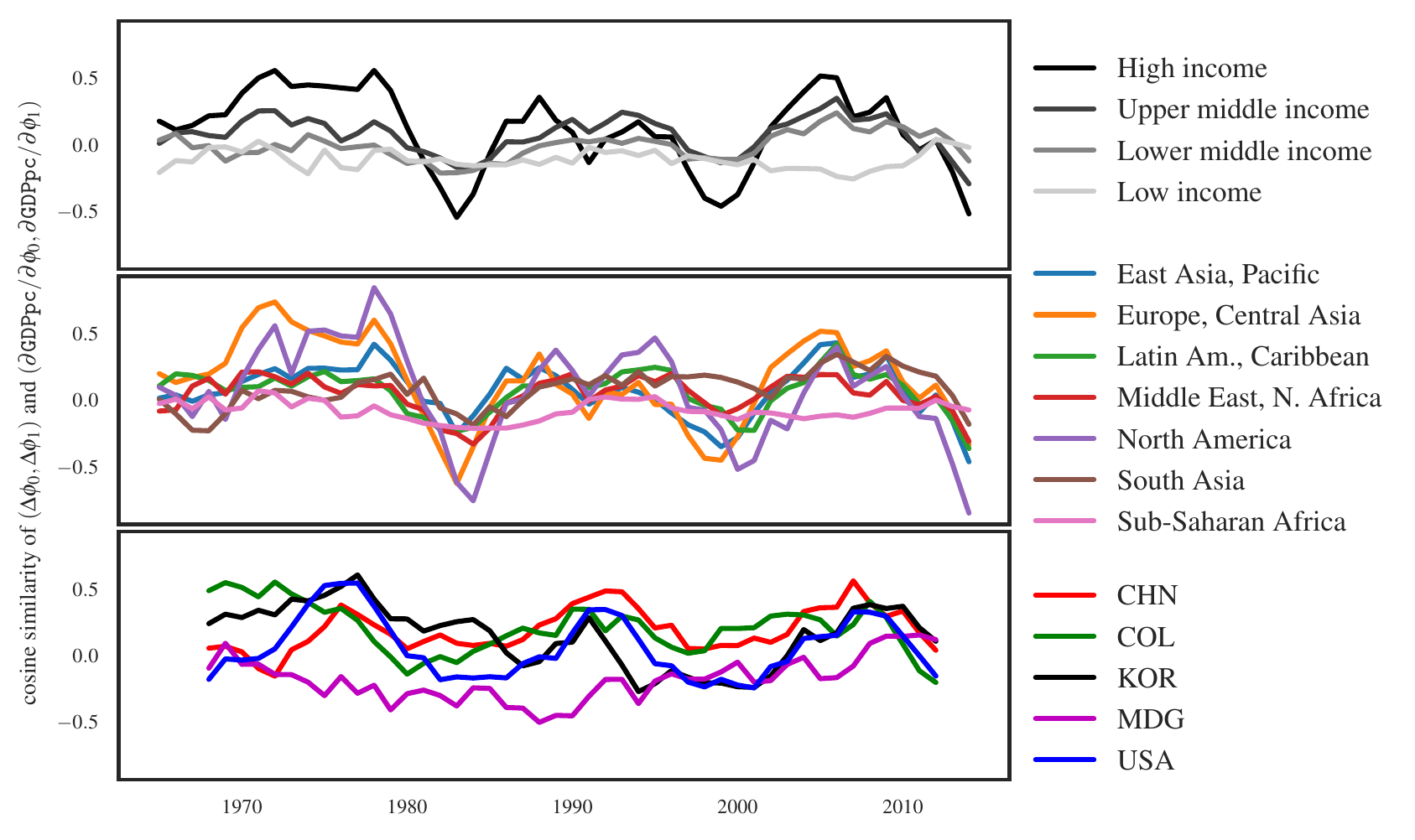}
\caption{
\textbf{Alignment of economies' changes in export baskets with the direction that would most increase per-capita incomes.}
Plotted is the cosine similarity between countries' movement in the first two principal components, $(\Delta \scoreFirstPC, \Delta \scoreSecondPC)$ and the gradient of the change in per-capita incomes with respect to the scores on the first two principal components, 
$
(\smoothTerm_{20}'(\scoreFirstPC), \smoothTerm_{21}'(\scoreSecondPC))$. 
A centered rolling average is applied to reduce noise (with window size $5$ in the first two rows and size $10$ in the third row). 
Income groups in the top row are from the World Bank.
}
\label{fig:cosine_similarity_by_income_and_region_and_country}
\end{center}
\end{figure}

Figure \ref{fig:cosine_similarity_by_income_and_region_and_country} shows the cosine similarity of countries' movement in $(\scoreFirstPC, \scoreSecondPC)$ and the gradient of how $\Delta \gdppc$ depends on $(\scoreFirstPC, \scoreSecondPC)$. By this measure, countries with higher incomes tend to be better hill climbers (i.e., high cosine similarity), and China has become an unusually good hill climber since the late $1990\text{s}$, while Madagascar has only recently moved slightly aligned with the gradient of per-capita incomes. 

\end{document}